\documentclass[10pt]{amsart}
\usepackage{amsmath}
\usepackage{amssymb}
\usepackage{amsfonts}
\usepackage{amsthm,amsxtra}
\usepackage{amsfonts}
\usepackage{nicefrac}

\theoremstyle{plain}

\newtheorem{corollary}{Corollary}[section]
\newtheorem{lemma}{Lemma}[section]
\newtheorem{proposition}{Proposition}[section]

\theoremstyle{definition}

\theoremstyle{remark}
\newtheorem{remark}{Remark}[section]

\newcommand{\PVI}{${\rm P}_{\rm VI}\:$}
\newcommand{\dPV}{${\rm dP}_{\rm V}\:$}

\newcommand{\ZZ}{\mathbb Z}

\newcommand{\CC}{\mathbb C}
\newcommand{\TT}{\mathbb T}

\newcommand{\shalf}{{\scriptstyle \frac{1}{2}}}
\newcommand{\half}{\nicefrac{1}{2}}
\newcommand{\thalf}{\nicefrac{3}{2}}

\begin{document}

\title[Discrete Painlev\'e equations, ...]
{\bf Discrete Painlev\'e equations for a class of \PVI $\tau$-functions given as 
$U(N)$ averages}

\author{P.J.~Forrester and N.S.~Witte}

\address{Department of Mathematics and Statistics,
University of Melbourne, Victoria 3010, Australia}
\email{\tt p.forrester@ms.unimelb.edu.au; n.witte@ms.unimelb.edu.au}

\begin{abstract}
In a recent work difference equations (Laguerre-Freud equations) for the 
bi-orthogonal polynomials and related quantities corresponding to the 
weight on the unit circle $ w(z)=\prod^m_{j=1}(z-z_j(t))^{\rho_j} $ were
derived.Here it is shown that in the case $ m=3 $ these difference equations,
when applied to the calculation of the underlying $U(N)$ average, reduce to a 
coupled system identifiable with that obtained by Adler and van Moerbeke using 
methods of the Toeplitz lattice and Virasoro constraints. Moreover it is shown 
that this coupled system can be reduced to yield the discrete fifth Painlev\'e
equation \dPV as it occurs in the theory of the sixth Painlev\'e system.
Methods based on affine Weyl group symmetries of B\"acklund transformations
have previously yielded the \dPV equation but with different parameters
for the same problem. We find the explicit mapping between the two forms.
Applications of our results are made to give recurrences for the gap 
probabilities and moments in the circular unitary ensemble of random matrices,
and to the diagonal spin-spin correlation function of the square lattice
Ising model. 
\end{abstract}

\maketitle

\section{Introduction}
\setcounter{equation}{0}

In a recent work \cite{FW_2004a} we undertook a study into differential and
difference structures associated with bi-orthogonal polynomials for the weight
on the unit circle
\begin{equation}
   w(z) = \prod^{m}_{j=1}(z-z_j(t))^{\rho_j}, \quad z=e^{i\theta}, \quad \theta\in(-\pi,\pi] .
\label{ops_gJwgt}
\end{equation} 
Our motivation was to eventually use these results to provide recurrences in 
$ N $  for the random matrix average
\begin{equation}
  \Big\langle \prod_{l=1}^N w(z_l) \Big\rangle_{U(N)} :=
  {1 \over (2 \pi )^N N!} \int^{\pi}_{-\pi}d\theta_1 \ldots \int^{\pi}_{-\pi}d\theta_N
  \prod^{N}_{l=1}w(z_l) \prod_{1\leq j<k \leq N}|z_k-z_j|^2 .
\label{ops_Uavge}
\end{equation}
Here $ U(N) $ denotes the unitary group of $ N\times N $ matrices, and the right
hand side exhibits the corresponding eigenvalue probability density function in the 
case of the Haar (uniform) measure.  

In this work we fulfill our original goal in the case $ m=3 $ by specialising the 
results of \cite{FW_2004a} as relevant to the calculation of (\ref{ops_Uavge}) with the
piecewise continuous version of (\ref{ops_gJwgt}) in the case $ m=3 $, 
\begin{equation}
   w(z) = t^{-\mu} z^{-\mu-\omega}(1+z)^{2\omega_1}(1+tz)^{2\mu}
   \begin{cases}
     1 & \theta \notin (\pi-\phi,\pi) \\
     1-\xi & \theta \in (\pi-\phi,\pi)
   \end{cases} ,
\label{VI_wgt}
\end{equation}
where $ \mu,\omega_1,\omega_2 $ are complex parameters ($ \omega=\omega_1+i\omega_2 $) 
and $ \xi, t=e^{i\phi} $ are complex variables ($ \phi\in [0,2\pi) $). 
By so studying (\ref{VI_wgt}) we are also completing a study begun
in \cite{FW_2003b} on recurrences satisfied by $ U(N) $ averages in random matrix 
theory from the viewpoint of orthogonal polynomial theory.

The average (\ref{ops_Uavge}) with weight (\ref{VI_wgt}) occurs in a variety of 
problems from mathematical physics. The identification of this average as a
$\tau$-function for the Painlev\'e VI system \cite{FW_2002b} has led to its
characterisation in terms of a solution of
the $\sigma$-form of the Painlev\'e VI equation. Regarding specific examples,
we first mention that the case $(\omega_1,\omega_2,\mu) = (0,0,0)$ is the 
generating function for the probability that the interval $(\pi-\phi,\pi)$ 
contains exactly $k$ eigenvalues in Dyson's circular unitary ensemble 
$ {\rm CUE}_N $ (which is equivalent to the unitary group with Haar measure). 
Similarly the case $(\omega_1,\omega_2,\mu) = (1,0,1)$ is (apart from a simple 
factor) the generating function for the probability density function of the event
that two eigenvalues from $ {\rm CUE}_{N+2} $ are an angle $\phi$ apart with 
exactly $k$ eigenvalues in between. The case $\xi = 2$, $\omega_2 = 0$, 
$\mu = \omega_1 = 1/2$ corresponds to the density matrix for the impenetrable 
Bose gas in periodic boundary conditions. It was studied in detail in 
\cite{FFGW_2002a}. Furthermore, in the case $\xi = 0$, one sees that (\ref{VI_wgt}) 
includes as special cases
\begin{align}
  & \Big\langle \prod_{l=1}^N z_l^{1/4} |1 + z_l |^{-1/2}
  (1 + k^{-2}z_l)^{1/2}\Big \rangle_{U(N)}, \qquad
  1/k^2 \le 1, \label{IM_Uavge}
  \\
  & \Big\langle \prod_{l=1}^N (1 + 1/z_l)^{v'}(1 + q^2 z_l)^v  
  \Big\rangle_{U(N)}, \qquad q^2 < 1. \label{IS_Uavge} 
\end{align}
The average (\ref{IM_Uavge}) is equivalent to the Toeplitz determinant given
by Onsager for the diagonal spin-spin correlation in the two-dimensional
Ising model \cite{McCW_1973}. As a PVI $\tau$-function it has been studied in 
\cite{JM_1980} and \cite{FW_2002b}. The average (\ref{IS_Uavge}) occurs as a cumulative
probability density in the study of processes relating to increasing
subsequences \cite{BR_2001,BO_2000}.

In the case $ \xi=0 $ of (\ref{VI_wgt}), the same problem as we are addressing here 
has been previously considered by Adler and van Moerbeke \cite{AvM_2002}, using 
methods of the Toeplitz lattice and Virasoro constraints. The difference equations 
obtained there were not identified with known integrable difference equations. 
Here we find that our formalism of bi-orthogonal polynomials leads to the very 
same difference equations. Moreover, we are able to show that they can be reduced
to examples of the discrete Painlev\'e V equation \dPV,
\begin{align}
  g_{n+1}g_n 
  & = t{(f_n+1-\alpha_2)(f_n+1-\alpha_0-\alpha_2) \over f_n(f_n+\alpha_3)} ,
  \label{dPV:a} \\
  f_n+f_{n-1} & = -\alpha_3 + {\alpha_1 \over g_n-1} +
                  {\alpha_4 t \over g_n-t} ,
  \label{dPV:b}
\end{align}
where 
$ \alpha_1 \mapsto \alpha_1+1,\alpha_2 \mapsto \alpha_2-1,\alpha_4 \mapsto \alpha_4+1 $
as $ n \mapsto n+1 $, fundamental in the theory of the \PVI system for its 
relationship to the B\"acklund transformations \cite{FW_2002b}, and its association
with the degeneration of the rational surface $ D^{(1)}_4 \to D^{(1)}_5 $ in the 
space of initial conditions \cite{Sa_2001}.

The \dPV system in relation to $ U(N) $ averages (specifically (\ref{IS_Uavge}))
was first found in the work of Borodin \cite{Bo_2001},\cite{BB_2002}. 
Subsequently the present authors \cite{FW_2003a} used methods based on the
affine Weyl group symmetries of B\"acklund transformations for the \PVI  system
to give \dPV recurrences for (\ref{ops_Uavge}) with weight (\ref{VI_wgt}).
These are different to the form of \dPV found from our transformation
of the recurrences from the bi-orthogonal polynomial theory, thus raising the
question as to the relationship between the two. This we answer by providing
the explicit transformation formulae.

A practical side of our work is that the reduced (in order) form of the Adler and 
van Moerbeke type recurrences are typically much simpler than their equivalent
form expressed in terms of \dPV. We make explicit these recurrences in the cases 
of the characteristic polynomial and gap probability for the $ {\rm CUE}_N $,
and the case of the diagonal spin-spin correlation function for the square lattice
Ising model.

In Section 2 general formulae from \cite{FW_2004a} required in subsequent sections 
are recalled. These formulae are used in Section 3 to derive $N$-recurrences in 
the case of weight (\ref{VI_wgt}), and furthermore a transformation to the \dPV 
system (\ref{dPV:a}), (\ref{dPV:b}) is given. The explicit transformation between 
the latter, and the \dPV system found in relation to (\ref{VI_wgt}) as a 
consequence of the Okamoto $\tau$-function theory \cite{FW_2003a}, is established
in Section 4. Application of the recurrences to random matrices and the Ising model 
is given in Section 5.

\section{Bi-orthogonal Polynomial Formalism}
\setcounter{equation}{0}

From the viewpoint of our work \cite{FW_2004a}, the weight (\ref{VI_wgt}) is a
particular example of the regular semi-classical class (\ref{ops_gJwgt}), characterised
by a special structure of their logarithmic derivatives
\begin{equation}
  \frac{1}{w(z)}\frac{d}{dz}w(z) = \frac{2V(z)}{W(z)}
  = \sum^m_{j=1}\frac{\rho_j}{z-z_j}, \quad \rho_j \in \CC.
\label{ops_scwgt2}
\end{equation}
Here $ V(z) $, $ W(z) $ are polynomials with $ {\rm deg}V(z) < m, {\rm deg}W(z)=m $.
We define bi-orthogonal polynomials $ \{\phi_n(z),\bar{\phi}_n(z)\}^{\infty}_{n=0} $
with respect to the weight $ w(z) $ on the unit circle by the orthogonality relation
\begin{equation}
  \int_{\TT} {d\zeta \over 2\pi i\zeta} w(\zeta)\phi_m(\zeta)\bar{\phi}_n(\bar{\zeta})
   = \delta_{m,n} ,
\label{ops_onorm}
\end{equation}
where $ \TT $ denotes the unit circle $ |\zeta|=1 $ with 
$ \zeta=e^{i\theta} $, $ \theta\in (-\pi,\pi] $. Notwithstanding the notation,
$ \bar{\phi}_n $ is not in general equal to the complex conjugate of $ \phi_n $.
We set
\begin{align}
  \phi_n(z)       &= \kappa_nz^n + l_nz^{n-1} + m_nz^{n-2} +\ldots +\phi_n(0) ,
  \nonumber\\
  \bar{\phi}_n(z) &= \kappa_nz^n + \bar{l}_nz^{n-1} + \bar{m}_nz^{n-2} +\ldots +\bar{\phi}_n(0) , 
  \nonumber
\end{align}
where again $ \bar{l}_n $, $ \bar{m}_n $, $ \bar{\phi}_n(0) $ are not in general 
equal to the corresponding complex conjugate. The coefficients are related by many 
coupled equations, two of the simplest being
\begin{equation}
   \kappa_n^2 = \kappa_{n-1}^2 + \phi_n(0)\bar{\phi}_n(0), \quad
   {l_{n} \over \kappa_{n}}-{l_{n-1} \over \kappa_{n-1}} = r_{n}\bar{r}_{n-1}.
   \label{ops_l}
\end{equation}

Denote the $ U(N) $ average (\ref{ops_Uavge}) by $ I_N[w] $. It is a basic fact 
that $ I_N[w] $ can also be written as the Toeplitz determinant
\begin{equation}
  I_n[w] = \det\left[ \int_{\TT}\frac{d\zeta}{2\pi i\zeta}w(\zeta)\zeta^{-j+k}
               \right]_{0\leq j,k\leq n-1} . 
\label{ops_Toep}
\end{equation}
With the so-called reflection coefficients specified by
\begin{equation}
  r_n = \frac{\phi_n(0)}{\kappa_n}, \quad 
  \bar{r}_n = \frac{\bar{\phi}_n(0)}{\kappa_n} ,
\end{equation}
it is a well known result in the theory of Toeplitz determinants that
\begin{equation}
   {I_{n+1}[w] I_{n-1}[w] \over (I_{n}[w])^2}
   = 1 - r_{n}\bar{r}_n .
\label{ops_I0}
\end{equation}
Introduce the reciprocal polynomial $ \phi^*_n(z) $ of the $ n$th degree polynomial
$ \bar{\phi}_n(z) $ by
\begin{equation}
  \phi^*_n(z) := z^n\bar{\phi}_n(1/z) .
\label{ops_recip}
\end{equation}
Fundamental to our study \cite{FW_2004a} is the matrix
\begin{equation}
   Y_n(z;t) := 
   \begin{pmatrix}
          \phi_n(z)   &  \epsilon_n(z)/w(z) \cr
          \phi^*_n(z) & -\epsilon^*_n(z)/w(z) \cr
   \end{pmatrix} ,
\label{ops_Ydefn}
\end{equation}
where
\begin{align}
   \epsilon_n(z)
   &:= \int_{\TT}{d\zeta \over 2\pi i\zeta}{\zeta+z \over \zeta-z}w(\zeta)
                   \phi_n(\zeta) ,
   \label{ops_eps:a} \\
   \epsilon^*_n(z)
   &:= -z^n\int_{\TT}{d\zeta \over 2\pi i\zeta}{\zeta+z \over \zeta-z}w(\zeta)
                   \bar{\phi}_n(\bar{\zeta}) .
   \label{ops_eps:b}
\end{align}
Thus we obtained the difference system
\begin{equation}
   Y_{n+1} := K_n Y_{n}
   = {1\over \kappa_n}
       \begin{pmatrix}
              \kappa_{n+1} z   & \phi_{n+1}(0) \cr
              \bar{\phi}_{n+1}(0) z & \kappa_{n+1} \cr
       \end{pmatrix} Y_{n} ,
\label{ops_Yrecur}
\end{equation}
and the differential system
\begin{multline}
   {d \over dz}Y_{n} := A_n Y_{n}
   \\
   = {1\over W(z)}
       \begin{pmatrix}
              -\left[ \Omega_n(z)+V(z)
                     -\dfrac{\kappa_{n+1}}{\kappa_n}z\Theta_n(z)
               \right]
            & \dfrac{\phi_{n+1}(0)}{\kappa_n}\Theta_n(z)
            \cr
              -\dfrac{\bar{\phi}_{n+1}(0)}{\kappa_n}z\Theta^*_n(z)
            &  \Omega^*_n(z)-V(z)
                     -\dfrac{\kappa_{n+1}}{\kappa_n}\Theta^*_n(z)
            \cr
       \end{pmatrix} Y_{n} .
\label{ops_YzDer}
\end{multline}

For regular semi-classical weights (\ref{ops_gJwgt}), the functions $ \Theta_n(z) $,
$ \Theta^*_n(z) $, $ \Omega_n(z) $ and $ \Omega^*_n(z) $ in (\ref{ops_YzDer}) are
polynomials of degree $ {\rm deg}\Omega_n(z)={\rm deg}\Omega^*_n(z)=m-1 $,
$ {\rm deg}\Theta_n(z)={\rm deg}\Theta^*_n(z)=m-2 $, independent of $ n $.
Explicitly, for large $z$
\begin{multline}
  \Theta_n(z) =
 (n+1+\sum^m_{j=1}\rho_j){\kappa_n \over \kappa_{n+1}}z^{m-2}
 \\
 + \bigg\{ -[(n+1+\sum^m_{j=1}\rho_j)\sum^m_{j=1}z_j - \sum^m_{j=1}\rho_j z_j]
             {\kappa_n \over \kappa_{n+1}}
            +(n+2+\sum^m_{j=1}\rho_j){\kappa^3_n \over \kappa^2_{n+1}\kappa_{n+2}}
             {\phi_{n+2}(0) \over \phi_{n+1}(0)}
 \\
 -(n+\sum^m_{j=1}\rho_j){\phi_{n+1}(0)\bar{\phi}_{n}(0) \over
                                   \kappa_{n+1}\kappa_{n}}
           -2{\kappa_nl_{n+1} \over \kappa^2_{n+1}}
   \bigg\}z^{m-3} + {\rm O}(z^{m-4}) ,
\label{ops_Thexp:a}
\end{multline}
\begin{multline}
  \Omega_n(z) =
 (1+\half\sum^m_{j=1}\rho_j)z^{m-1}
 \\
 + \bigg\{ -\half(\sum^m_{j=1}\rho_j)(\sum^m_{j=1}z_j)
            +\half \sum^m_{j=1}\rho_j z_j
            -\sum^m_{j=1}z_j
 \\
 + (n+2+\sum^m_{j=1}\rho_j)
           {\kappa^2_n \over \kappa_{n+2}\kappa_{n+1}}
           {\phi_{n+2}(0) \over \phi_{n+1}(0)}
     - {l_{n+1} \over \kappa_{n+1}} \bigg\}z^{m-2} + {\rm O}(z^{m-3}) .
\label{ops_Omexp:a}
\end{multline}
Also obtained in \cite{FW_2004a}, and relevant to the present study, are the 
small $z$ expansions
\begin{multline}
 \Theta_n(z) =
 [2V(0)-nW'(0)]{\phi_{n}(0) \over \phi_{n+1}(0)}
 \\
 + \bigg\{ [2V'(0)-\shalf nW''(0)]{\phi_{n}(0) \over \phi_{n+1}(0)}
          + [2V(0)-(n-1)W'(0)]{\kappa_n\phi_{n-1}(0) \over \kappa_{n-1}\phi_{n+1}(0)}
 \\
 + \left( [(n+1)W'(0)-2V(0)]{\bar{l}_{n+1} \over \kappa_{n+1}}
                   -[(n-1)W'(0)-2V(0)]{\bar{l}_{n-1} \over \kappa_{n+1}} \right)
           {\phi_{n}(0) \over \phi_{n+1}(0)}
   \bigg\}z
 \\ + {\rm O}(z^{2}) ,
\label{ops_Thexp:b}
\end{multline}
\begin{multline}
  \Omega_n(z) =
 V(0)-nW'(0)
 \\
 + \bigg\{ V'(0)-\shalf nW''(0)
           +\left( V(0){\kappa_n \over \kappa_{n+1}}
                    +[V(0)-nW'(0)]{\kappa_{n+1} \over \kappa_{n}} \right)
            {\phi_{n}(0) \over \phi_{n+1}(0)}
 \\
 + [V(0)-nW'(0)]{\bar{l}_n \over \kappa_n}
   - [V(0)-(n+1)W'(0)]{\bar{l}_{n+1} \over \kappa_{n+1}} \bigg\}z + {\rm O}(z^2) .
\label{ops_Omexp:b}
\end{multline}
Analogous formulae hold for $ \Theta^*_n(z) $ and $ \Omega^*_n(z) $
(these can be found in \cite{FW_2004a}).

The coefficient functions $ \Omega_n(z),\Omega^*_n(z), \Theta_n(z), \Theta^*_n(z) $
satisfy sets of difference and functional relations given in \cite{FW_2004a}.
Relevant for the present study are the coupled equations
\begin{multline}
   \left( {\phi_{n+1}(0) \over \phi_{n}(0)}+{\kappa_{n+1} \over \kappa_{n}}z
   \right) (\Omega_{n-1}(z) - \Omega_{n}(z))
   \\
   + {\kappa_{n}\phi_{n+2}(0) \over \kappa_{n+1}\phi_{n+1}(0)}z\Theta_{n+1}(z)
   - {\kappa_{n-1}\phi_{n+1}(0) \over \kappa_{n}\phi_{n}(0)}z\Theta_{n-1}(z)
   - {\phi_{n+1}(0) \over \phi_{n}(0)}{W(z) \over z} = 0 ,
\label{ops_rrCf:b}
\end{multline}
\begin{multline}
   \left( {\kappa_{n+1} \over \kappa_{n}}+{\bar{\phi}_{n+1}(0) \over \bar{\phi}_{n}(0)}z
   \right) (\Omega^*_{n-1}(z) - \Omega^*_{n}(z))
   \\
   + {\kappa_{n}\bar{\phi}_{n+2}(0) \over \kappa_{n+1}\bar{\phi}_{n+1}(0)}
      z\Theta^*_{n+1}(z)
   - {\kappa_{n-1}\bar{\phi}_{n+1}(0) \over \kappa_{n}\bar{\phi}_{n}(0)}
      z\Theta^*_{n-1}(z)
   + {\kappa_{n+1} \over \kappa_{n}}{W(z) \over z} = 0 ,
\label{ops_rrCf:d}
\end{multline}
\begin{multline}
  \Omega^*_{n+1}(z) + \Omega_{n}(z)
  - \left( {\kappa_{n+2} \over \kappa_{n+1}}
           +{\bar{\phi}_{n+2}(0) \over \bar{\phi}_{n+1}(0)}z
    \right)\Theta^*_{n+1}(z) \\
  - {\kappa_{n+1} \over \kappa_{n}}(z\Theta_n(z)-\Theta^*_n(z)) - {W(z) \over z}= 0 ,
\label{ops_rrCf:g}
\end{multline}
\begin{multline}
   \Omega^*_{n}(z) - \Omega^*_{n+1}(z)
   + {\kappa_{n+2} \over \kappa_{n+1}}
    \left( 1+{\phi_{n+1}(0) \over \kappa_{n+1}}{\bar{\phi}_{n+2}(0) \over \kappa_{n+2}}z
    \right)\Theta^*_{n+1}(z) \\
   + {\phi_{n+1}(0)\bar{\phi}_{n+1}(0) \over \kappa_{n+1}\kappa_{n}}z\Theta_n(z)
   - {\kappa_{n+1} \over \kappa_{n}}\Theta^*_{n}(z) = 0 ,
\label{ops_rrCf:h}
\end{multline}
\begin{gather}
  {\phi_{n+1}(0) \over \phi_{n}(0)}\Theta_{n}(z)
 - {\kappa_{n} \over \kappa_{n-1}}z\Theta_{n-1}(z)
 = {\bar{\phi}_{n+1}(0) \over \bar{\phi}_{n}(0)}z\Theta^*_n(z)
  - {\kappa_{n} \over \kappa_{n-1}} \Theta^*_{n-1}(z) ,
 \label{ops_rrCf:i} \\
  \Omega^*_{n}(z)-\Omega_{n}(z)
  = - {\kappa_{n+1} \over \kappa_{n}}(z\Theta_{n}(z)-\Theta^*_n(z))+n{W(z) \over z} ,
 \label{ops_rrCf:j}
\end{gather}
\begin{equation}
  \Omega^*_{n}(z)+\Omega_{n}(z)
  = {\kappa^2_{n} \over \kappa^2_{n+1}}
    \left[{\phi_{n+2}(0) \over \phi_{n+1}(0)}\Theta_{n+1}(z)
          + {\kappa_{n+1} \over \kappa_{n}} \Theta^*_{n}(z)\right]
    + {W(z) \over z} .
 \label{ops_rrCf:k}
\end{equation}
In addition to the above coupled equations, evaluations of the coefficient functions 
at the singular points satisfy bilinear relations \cite{FW_2004a}
\begin{gather}
  \Omega^2_n(z_j)
  = {\kappa_n \phi_{n+2}(0) \over \kappa_{n+1} \phi_{n+1}(0)}z_j
      \Theta_n(z_j)\Theta_{n+1}(z_j)+V^2(z_j) ,
  \label{ops_OTeq:a} \\
  \Omega^{*2}_n(z_j)
  = {\kappa_n \bar{\phi}_{n+2}(0) \over \kappa_{n+1} \bar{\phi}_{n+1}(0)}z_j
      \Theta^*_n(z_j)\Theta^*_{n+1}(z_j)+V^2(z_j) ,
  \label{ops_OTeq:b} \\
  \left[\Omega_{n-1}(z_j)
    -{\kappa^2_{n-1} \over \kappa^2_{n}}{\phi_{n+1}(0) \over \phi_{n}(0)}\Theta_n(z_j)
  \right]^2
  = {\phi_{n+1}(0)\bar{\phi}_{n}(0) \over \kappa^2_{n}}
      \Theta_n(z_j)\Theta^*_{n-1}(z_j)+V^2(z_j) ,
  \label{ops_OTeq:c}
\end{gather}
\begin{multline}
  \left[\Omega^*_{n-1}(z_j)
    -{\kappa^2_{n-1} \over \kappa^2_{n}}{\bar{\phi}_{n+1}(0) \over \bar{\phi}_{n}(0)}
      z_j\Theta^*_n(z_j)
  \right]^2
  \\
  = {\kappa_{n-1}\bar{\phi}_{n+1}(0)\phi_{n}(0) \over \kappa^3_{n}}
      z^2_j\Theta^*_n(z_j)\Theta_{n-1}(z_j)+V^2(z_j) ,
  \label{ops_OTeq:d}
\end{multline}
\begin{gather}
    {\phi_{n+1}(0)\bar{\phi}_{n+1}(0) \over \kappa^2_{n}}
     z_j\Theta_n(z_j)\Theta^*_{n}(z_j)+V^2(z_j)
  = \left[\Omega_{n}(z_j)-{\kappa_{n+1} \over \kappa_{n}}z_j\Theta_n(z_j)\right]^2 ,
  \label{ops_OTeq:e} \\
  \phantom{
    {\phi_{n+1}(0)\bar{\phi}_{n+1}(0) \over \kappa^2_{n}}
      \Theta_n(z_j)\Theta^*_{n}(z_j)+V^2(z_j) }
  = \left[\Omega^*_{n}(z_j)-{\kappa_{n+1} \over \kappa_{n}}\Theta^*_n(z_j)\right]^2 .
  \label{ops_OTeq:f}
\end{gather}
It is these relations that lead directly to one of the pair of coupled discrete 
Painlev\'e equations. Deformation derivatives of the system (\ref{ops_Ydefn})
with respect to arbitrary trajectories of the singularities $ z_j(t) $ are given 
in \cite{FW_2004a}, leading to the Schlesinger equations from the theory of
isomonodromic deformations of linear systems. The particular result that we require 
here is
\begin{align}
   \frac{1}{r_n}\frac{d}{dt}r_n & =
   \half\sum^m_{j=1}\rho_j\frac{1}{z_j}\frac{d}{dt}z_j 
   {\Omega_{n-1}(z_j)-V(z_j) \over V(z_j)} ,
   \label{ops_rdot} \\
   \frac{1}{\bar{r}_n}\frac{d}{dt}\bar{r}_n  & =
   \half\sum^m_{j=1}\rho_j\frac{1}{z_j}\frac{d}{dt}z_j
   {\Omega^*_{n-1}(z_j)+V(z_j) \over V(z_j)} .
   \label{ops_rCdot}
\end{align}

\section{The Semi-classical Class $ m=3 $ and \dPV}
\label{PVIsection}
\setcounter{equation}{0}

\subsection{Coupled Recurrences for the Reflection Coefficients}

Here we will consider the general theory from \cite{FW_2004a} revised above,
applied to the simplest instance of the semi-classical weight (\ref{ops_gJwgt}), 
namely $ m=3 $ singular points with two fixed at $ z=0,-1 $ and the third
variable at $ z=-1/t $. 
Explicitly we consider the unitary group average (\ref{ops_Uavge}) in the case of
the weight (\ref{VI_wgt}). For this to make immediate sense we require $ t \in \TT $,
but it can analytically continued off the unit circle. 
Note that when $ t \in \TT $, $ \mu, \omega_1, \omega_2, \xi \in \mathbb{R} $
and $ \xi < 1 $ the weight (\ref{VI_wgt}) is real and positive. The corresponding
Toeplitz determinant (\ref{ops_Toep}) is then hermitian and as a consequence 
$ \bar{r}_n $ is the complex conjugate of $ r_n $, but generally this is not the case.
For parameters such that $ \Re(\mu), \Re(\omega_1) > -1/2 $, $ t \in \TT $
the Toeplitz matrix elements, $ w_{j-k} $ say, can be evaluated in terms of the 
Gauss hypergeometric function and so analytically continued into the general 
parameter space. There are several forms for this, which exhibit the analytic
structure about the special points $ t = 0,1, \infty $. We will make note of two 
such expansions, relating to the special points $ t=0 $ and $ t=1 $.

\begin{lemma}
For all points in parameter space that the functions below have meaning, the 
analytic continuation of the Toeplitz matrix elements $ w_n $ for the weight 
(\ref{VI_wgt}) is given by
\begin{multline}
   t^{\mu}w_{n} = {\Gamma(2\omega_1+1) \over  
                    \Gamma(1+n+\mu+\omega)\Gamma(1-n-\mu+\bar{\omega})}
   {}_2F_1(-2\mu,-n-\mu-\omega;1-n-\mu+\bar{\omega};t) \\
 + {\xi \over 2\pi i} e^{\pm \pi i(n+\mu-\bar{\omega})}
    {\Gamma(2\mu+1)\Gamma(2\omega_1+1) \over \Gamma(2\mu+2\omega_1+2)}
    t^{n+\mu-\bar{\omega}}(1-t)^{2\mu+2\omega_1+1}
 \\
   \times {}_2F_1(2\mu+1,1+n+\mu+\omega;2\mu+2\omega_1+2;1-t) ,
\label{VI_toepM}
\end{multline}
where the $ \pm $ sign is taken accordingly as $ {\rm Im}(t) \gtrless 0 $.
This can also be written as
\begin{multline}
   t^{\mu}w_{n} = 
   \left\{ 1+\xi{e^{\pm\pi i(n+\mu-\bar{\omega})} \over 
           2i\sin\pi(n+\mu-\bar{\omega})}
   \right\}
   {\Gamma(2\omega_1+1) \over
                    \Gamma(1+n+\mu+\omega)\Gamma(1-n-\mu+\bar{\omega})} \\
   \times{}_2F_1(-2\mu,-n-\mu-\omega;1-n-\mu+\bar{\omega};t) \\
   -\xi{e^{\pm\pi i(n+\mu-\bar{\omega})} \over 2i\sin\pi(n+\mu-\bar{\omega})}
       {\Gamma(2\mu+1) \over
                    \Gamma(1+n+\mu-\bar{\omega})\Gamma(1-n+\mu+\bar{\omega})} \\
           \times t^{n+\mu-\bar{\omega}}(1-t)^{2\mu+2\omega_1+1}                               
   {}_2F_1(2\mu+1,1+n+\mu+\omega;1+n+\mu-\bar{\omega};t).
\label{VI_toepM2}
\end{multline}
\end{lemma}
\begin{proof}
This follows from the generalisation of the Euler integral for the Gauss hypergeometric
function and consideration of the consistent phases for the branch cuts linking the
singular points, see \cite{Kl_1933} pp. 91, section 17 "Verallgemeinerung der Eulersche
Integrale".
\end{proof}

\begin{remark}
The first of these forms was given in \cite{FW_2003a}.
\end{remark}

For the description of (\ref{VI_wgt}) in terms of the characterisation 
(\ref{ops_scwgt2}) of general semi-classical weights we have
\begin{gather}
  m=3 , \quad \{z_j\}^{3}_{j=1} = \{0,-1,-1/t\}, \quad 
  \{\rho_j\}^{3}_{j=1} = \{-\mu-\omega,2\omega_1,2\mu\}, \\
  2V(0) = -(\mu+\omega)t^{-1}, \; W'(0) = t^{-1}, \quad 
   V(-t^{-1}) = \mu{1-t \over t^2} .
\end{gather}
Also, according to (\ref{ops_Thexp:a})-(\ref{ops_Omexp:b}), the explicit forms of
the coefficient functions in this case is
\begin{gather}
\begin{split}
  \Theta_N(z)
 = & {\kappa_N \over \kappa_{N+1}}\left[ (N+1+\mu+\bar{\omega})z
       -{r_N \over r_{N+1}}(N+\mu+\omega)t^{-1} \right] ,
\end{split}
 \label{VI_theta:a} \\
\begin{split}
  \Theta^*_N(z)
 = & {\kappa_N \over \kappa_{N+1}}\left[
       -{\bar{r}_N \over \bar{r}_{N+1}}(N+\mu+\bar{\omega})z
       +(N+1+\mu+\omega)t^{-1} \right] ,
\end{split}
 \label{VI_theta:b}
\end{gather}
\begin{multline}
   \Omega_N(z)
   = [1+\shalf(\mu+\bar{\omega})]z^2
   \\
   + \left\{ (N+2+\mu+\bar{\omega})(1-r_{N+1}\bar{r}_{N+1}){r_{N+2} \over r_{N+1}}
             -{l_{N+1} \over \kappa_{N+1}} \right.
   \\
     \left.  +[1+\shalf (\mu+\bar{\omega})]{1+t \over t}-\omega_1-{\mu \over t} \right\}z
    - [N+\shalf (\mu+\omega)]t^{-1} ,
 \label{VI_omega:a}
\end{multline}
\begin{multline}
   \Omega^*_N(z)
   = -\shalf (\mu+\bar{\omega})z^2
   \\
   + \left\{ {l_{N+1} \over \kappa_{N+1}}-(N+\mu+\bar{\omega})
             (1-r_{N+1}\bar{r}_{N+1}){\bar{r}_{N} \over \bar{r}_{N+1}}
             -\shalf (\mu+\bar{\omega}){1+t \over t}+\omega_1+{\mu \over t} \right\}z
   \\
   +[N+1+\shalf (\mu+\omega)]t^{-1} .
 \label{VI_omega:b}
\end{multline}

We are seeking a closed system of recurrences for $ r_N $ and $ \bar{r}_N $, since 
according to (\ref{ops_I0}) these quantities determine the Toeplitz determinant,
or equivalently the $ U(N) $ average. One such recurrence is quite straight forward.
\begin{lemma}
The reflection coefficients for the weight (\ref{VI_wgt}) satisfy the homogeneous
second-order difference equation
\begin{multline}
   (N+1+\mu+\bar{\omega})tr_{N+1}\bar{r}_{N} - (N-1+\mu+\bar{\omega})tr_{N}\bar{r}_{N-1}
  \\ = (N+1+\mu+\omega)\bar{r}_{N+1}r_{N} - (N-1+\mu+\omega)\bar{r}_{N}r_{N-1} .
\label{VI_2ndRR}
\end{multline}
\end{lemma}

\begin{proof}
This equation can be found immediately from the
general theory of Section 2 in many ways. By equating coefficients of $ z $
in the functional-difference equation (\ref{ops_rrCf:g}) using (\ref{VI_theta:a},
\ref{VI_theta:b},\ref{VI_omega:a},\ref{VI_omega:b}), all are trivially satisfied
except for the $ z $ coefficient, which is precisely (\ref{VI_2ndRR}). Similarly
starting with (\ref{ops_rrCf:h}) and employing (\ref{VI_theta:a},
\ref{VI_theta:b},\ref{VI_omega:b}), one finds (\ref{VI_2ndRR}). Alternatively 
one could start with either (\ref{ops_rrCf:i}) or (\ref{ops_rrCf:k}) and arrive at the
same result
\end{proof}

We will use this result in the derivation of a sequence of lemmas, which lead us to 
the sought closed system of coupled recurrences.
\begin{corollary}
The sub-leading coefficients $ l_{N}, \bar{l}_{N} $ satisfy the linear inhomogeneous 
equation
\begin{equation}
   (N+\mu+\bar{\omega})tl_{N} - (N+\mu+\omega)\bar{l}_{N}
   = N\left[ \mu(t-1)+\bar{\omega}-\omega t \right]\kappa_{N} .
\label{VI_lRecur}
\end{equation}
\end{corollary}

\begin{proof}
By substituting the general expression for the first difference of $ l_N, \bar{l}_{N} $ 
using (\ref{ops_l}) in (\ref{VI_2ndRR}) one finds that it can be summed exactly to yield
\begin{multline}
  (N+1+\mu+\bar{\omega})t{l_{N+1} \over \kappa_{N+1}}
  - (N+1+\mu+\omega){\bar{l}_{N+1} \over \kappa_{N+1}} \\
  - (N+\mu+\bar{\omega})t{l_{N} \over \kappa_{N}}
  + (N+\mu+\omega){\bar{l}_{N} \over \kappa_{N}}
  = \mu(t-1)+\bar{\omega}-\omega t .
\label{VI_lRecurD}
\end{multline}
This can be summed once more to yield the stated result.
\end{proof}

\begin{remark}
One could alternatively proceed via the Freud approach \cite{Fr_1976} (see also
\cite{FW_2003b}) and consider the integral
\begin{equation}
    \int_{\TT} {dz \over 2\pi iz} (1+z)(1+tz)
  \left[ -{\mu+\omega \over z} + {2\omega_1 \over 1+z} + {2\mu t\over 1+tz} \right]
   w(z) \phi_{N}(z)\bar{\phi}_{N}(\bar{z}).
\end{equation}
Here we recognise the logarithmic derivative of the weight function in the 
integrand
\begin{equation}
   {w' \over w} = -{\mu+\omega \over z} + {2\omega_1 \over 1+z} + {2\mu t\over 1+tz},
\end{equation}
and by evaluating the integral in the two ways we find a linear equation
for $ l_{N} $, namely (\ref{VI_lRecurD}).
\end{remark}

\begin{lemma}
The sub-leading coefficients are related to the reflection coefficients by
\begin{gather}
  \bar{l}_{N}/\kappa_{N} + tl_{N}/\kappa_{N} -N(t+1)
  \nonumber \\
  = {1-r_{N}\bar{r}_{N} \over r_{N}}
    \left[ (N+1+\mu+\bar{\omega})tr_{N+1} + (N-1+\mu+\omega)r_{N-1} \right] ,
  \label{VI_magnus:a} \\
  = {1-r_{N}\bar{r}_{N} \over \bar{r}_{N}}
    \left[ (N+1+\mu+\omega)\bar{r}_{N+1} + (N-1+\mu+\bar{\omega})t\bar{r}_{N-1} \right] .
  \label{VI_magnus:b}
\end{gather}
\end{lemma}

\begin{proof}
The first relation follows from a comparison of the coefficients of $ z $ for 
$ \Omega_N(z) $ given the two distinct expansions, the first by (\ref{ops_Omexp:a}) 
which reduces to (\ref{VI_omega:a}) and the second by the specialisation of 
(\ref{ops_Omexp:b}). The second relation follows from identical arguments applied to
$ \Omega^*_N(z) $ or by employing (\ref{VI_2ndRR}) in the first relation.
\end{proof}

\begin{remark}
The first result appears in the Magnus derivation \cite{Ma_2000} for the generalised
Jacobi weight, with $ \theta_1 = \pi-\phi, \theta_2 = \pi $, 
$ \alpha = \mu, \beta = \omega_1, \gamma = -\omega_2 $. Then Equation (14) of that work 
is precisely (\ref{VI_magnus:a}).
\end{remark}

\begin{remark}
The Magnus relation (\ref{VI_magnus:a}) can also be found by employing the Freud 
method. In this one uses integration by parts on the integral
\begin{equation}
   \int_{\TT} {dz \over 2\pi iz} z^{-1}(1+z)(1+tz)
   w'(z) \phi_{N+1}(z)\bar{\phi}_{N}(\bar{z}),
\end{equation}
and in the term involving $ \phi'_{N+1}(z) $ one employs (\ref{ops_YzDer})
for the derivative and (\ref{VI_theta:a}), (\ref{VI_omega:a}) for the coefficient
functions. Equating this expression to a direct evaluation of the integral then 
yields (\ref{VI_magnus:a}). 
\end{remark}

\begin{lemma}
The sub-leading coefficient $ l_N $ can be expressed in terms of the reflection 
coefficients in the following ways
\begin{align}\label{VI_lSoln:a}
   2t{l_N \over \kappa_{N}} 
  = & (N+1+\mu+\bar{\omega})t({r_{N+1} \over r_{N}}-r_{N+1}\bar{r}_{N})
  +(N-1+\mu+\omega){r_{N-1} \over r_{N}}
  \\
    & -(N-1+\mu+\bar{\omega})r_{N}\bar{r}_{N-1} + (N+\mu-\omega)t + N-\mu+\bar{\omega} ,
  \nonumber \\
  = & (N+1+\mu+\omega){\bar{r}_{N+1} \over \bar{r}_{N}}
   +(N-1+\mu+\bar{\omega})t({\bar{r}_{N-1} \over \bar{r}_{N}}-r_{N}\bar{r}_{N-1})
  \label{VI_lSoln:b} \\
    & -(N+1+\mu+\bar{\omega})tr_{N+1}\bar{r}_{N} + (N+\mu-\omega)t + N-\mu+\bar{\omega} ,
  \nonumber
\end{align} 
as well as analogous expressions for $ \bar{l}_N $.
\end{lemma}

\begin{proof}
The first expression follows from a comparison of the $ z^0 $ coefficients for 
$ \Theta_N(z) $ evaluated using both (\ref{ops_Thexp:a}) and (\ref{ops_Thexp:b}).
The second relation follows from an applying the same reasoning to $ \Theta^*_N(z) $.
\end{proof}

We will refer to the order of a system of coupled difference equations with two 
variables, $ r_n, \bar{r}_n $ say, as $ q/p $ where $ q\in \mathbb{Z}_{\geq 0} $ 
refers to the order of $ r_n $ and $ p\in \mathbb{Z}_{\geq 0} $ refers to the 
order of $ \bar{r}_n $.

\begin{corollary}
The reflection coefficients of the OPS for the weight (\ref{VI_wgt}) satisfy the
$ 2/2 $ order recurrence relations
\begin{gather}
   tr_{N}\bar{r}_{N-1} + r_{N-1}\bar{r}_{N} -t-1
   \nonumber \\
   = {1-r_{N}\bar{r}_{N} \over r_{N}}
       \left[ (N+1+\mu+\bar{\omega})tr_{N+1} + (N-1+\mu+\omega)r_{N-1} \right]
   \nonumber\\
   \quad
     - {1-r_{N-1}\bar{r}_{N-1} \over \bar{r}_{N-1}}
       \left[ (N+\mu+\omega)\bar{r}_{N} + (N-2+\mu+\bar{\omega})t\bar{r}_{N-2} 
       \right] ,
   \label{VI_rRecur:a}
   \\
   = {1-r_{N}\bar{r}_{N} \over \bar{r}_{N}}
       \left[ (N+1+\mu+\omega)\bar{r}_{N+1} + (N-1+\mu+\bar{\omega})t\bar{r}_{N-1}
       \right]
   \nonumber\\
   \quad
     - {1-r_{N-1}\bar{r}_{N-1} \over r_{N-1}}
       \left[ (N+\mu+\bar{\omega})tr_{N} + (N-2+\mu+\omega)r_{N-2} \right] ,
   \label{VI_rRecur:b}
\end{gather}
and those specifying the solution for (\ref{ops_I0}) have the initial values
\begin{equation} 
   r_{0} = \bar{r}_{0} = 1, \quad
   r_{1} = -w_{-1}/w_0, \quad \bar{r}_{1} = -w_{1}/w_{0},
   \label{VI_rInitial}
\end{equation} 
where the Toeplitz matrix elements are given in (\ref{VI_toepM2}).
\end{corollary}

\begin{proof}
Solving (\ref{VI_magnus:a}) for the combination of $ l_{N}, \bar{l}_{N} $ and 
differencing this, one arrives at (\ref{VI_rRecur:a}). This however is of order 
$ 3/1 $ but by employing (\ref{VI_2ndRR}) we can reduce the order in $ r_{N} $ of 
the recurrence to second order. The other member of the pair (\ref{VI_rRecur:b}) 
is found in an identical manner starting with 
(\ref{VI_magnus:b}).
\end{proof}

\begin{remark}
The second-order difference (\ref{VI_rRecur:a}) also follows immediately from 
equating
the polynomials in $ z $ arising in the functional-difference (\ref{ops_rrCf:b}), 
after employing (\ref{VI_theta:a},\ref{VI_omega:a}). The other member of the pair, 
(\ref{VI_rRecur:b}), follows from the functional-difference (\ref{ops_rrCf:d}),
after using (\ref{VI_theta:b},\ref{VI_omega:b}).
\end{remark}

\subsection{Relationship to recurrences of Adler and van Moerbeke}

In their most general example, Adler and van Moerbeke \cite{AvM_2002} also 
considered recurrences from the $ U(N) $ average with weight (\ref{VI_wgt})
($ \xi=0 $ case). Their study proceeded via the viewpoint of the Toeplitz lattice
and Virasoro constraints. In terms of their variables we should set 
$ P_1 = P_2 = 0, d_1 = t^{-1/2}, d_2 = t^{1/2} $, and without loss of generality
$ \gamma''_1 = \gamma'_2 = 0 $. For the other parameters 
$ \gamma = \mu-\omega, \gamma'_1 = 2\omega_1, \gamma''_2 = 2\mu $.
There is a slight difference in the dependent variables due to the additional
factor of $ t $, so that we have the identification $ x_{N} = (-1)^Nt^{N/2}r_{N} $,
$ y_{N} = (-1)^Nt^{-N/2}\bar{r}_{N} $ and $ v_{N} = 1-r_{N}\bar{r}_{N} $. 
Generalising their working one finds that their Equation (0.0.14) implies
\begin{multline}
   -(N+1+\mu+\bar{\omega})x_{N+1}y_{N} + (N+1+\mu+\omega)x_{N}y_{N+1} \\
   +(N-1+\mu+\bar{\omega})x_{N}y_{N-1} - (N-1+\mu+\omega)x_{N-1}y_{N} = 0.
\end{multline}
Now by transforming to our $ r_{N},\bar{r}_{N} $ and employing (\ref{ops_l})
one finds this is precisely (\ref{VI_2ndRR}), which we showed is solved by
(\ref{VI_lRecur}). Their inhomogeneous Equation (0.0.15) now takes the form
\begin{multline}
   -v_{N}\left[ (N+1+\mu+\bar{\omega})x_{N+1}y_{N-1}+N+\mu+\omega \right] \\
   +v_{N-1}\left[ (N-2+\mu+\bar{\omega})x_{N}y_{N-2}+N-1+\mu+\omega \right] \\
   +x_{N}y_{N-1}(x_{N}y_{N-1}+t^{1/2}+t^{-1/2}) \\
  = -v_{1}\left[ (2+\mu+\bar{\omega})x_{2}+1+\mu+\omega \right]
      +x_{1}(x_{1}+t^{1/2}+t^{-1/2}) .
\end{multline}
Upon recasting this into our variables and manipulating, it then becomes
\begin{multline}
   tr_{N}\bar{r}_{N-1} + \bar{r}_{N}r_{N-1}-t-1
   \nonumber \\
   - {1-r_{N}\bar{r}_{N} \over r_{N}}
       \left[ (N+1+\mu+\bar{\omega})tr_{N+1} + (N-1+\mu+\omega)r_{N-1} \right] \\
   \quad
   + {1-r_{N-1}\bar{r}_{N-1} \over \bar{r}_{N-1}}
     \left[ (N+\mu+\omega)\bar{r}_{N} + (N-2+\mu+\bar{\omega})t\bar{r}_{N-2} \right] \\
   = {1-(1-r_{1}\bar{r}_{1})\left[(2+\mu+\bar{\omega})tr_{2}+1+\mu+\omega \right]
             +r_{1}(tr_{1}-t-1) \over r_{N}\bar{r}_{N-1}} .
\end{multline}
However using the identity
\begin{multline*}
  c{}_2F_1(a,b;c;x) 
  \\
  = [c+(1+b-a)x]{}_2F_1(a,b+1;c+1;x) -\frac{b+1}{c+1}(1+c-a)x{}_2F_1(a,b+2;c+2;x) ,
\end{multline*}
we note that the right-hand side is identically zero for the initial conditions 
(\ref{VI_rInitial}) and the recurrence is not genuinely inhomogeneous, thus 
yielding our first relation above, (\ref{VI_rRecur:a}).

\subsection{Transformation of the recurrences to \dPV}

We seek recurrences for $ r_N, \bar{r}_N $ which are of the form of the discrete
Painlev\'e system (\ref{dPV:a}), (\ref{dPV:b}). For this purpose a number of
distinct forms of the former will be presented.

\begin{proposition}
The reflection coefficients satisfy a system of a $ 2/0 $ order recurrence relation
\begin{multline}\label{VI_2+0rRecur:a}
  \Big\{ (1-r_N\bar{r}_N) \left[
  (N+1+\mu+\bar{\omega})(N+\mu+\bar{\omega})tr_{N+1} \right.
  \\ \left. - (N+\mu+\omega)(N-1+\mu+\omega)r_{N-1} \right]
     + N(N+2\omega_1)(t-1)r_N \Big\}
  \\ \times
  \Big\{ (1-r_N\bar{r}_N) \left[
  (N+1+\mu+\bar{\omega})(N+\mu+\bar{\omega})tr_{N+1} \right.
  \\ \left. - (N+\mu+\omega)(N-1+\mu+\omega)r_{N-1} \right] 
     + (N+2\mu)(N+2\mu+2\omega_1)(t-1)r_N \Big\}
  \\ 
  = -(2N+2\mu+2\omega_1)^2t(1-r_N\bar{r}_N)
  \\ \times
  \left[ (N+1+\mu+\bar{\omega})r_{N+1}+(N+\mu+\omega)r_{N} \right]
  \\ \times
  \left[ (N+\mu+\bar{\omega})r_{N}+(N-1+\mu+\omega)r_{N-1} \right] ,
\end{multline}
and a $ 0/2 $ order recurrence relation which is just (\ref{VI_2+0rRecur:a}) with the 
replacements $ \omega \leftrightarrow \bar{\omega} $ and 
$ t^{\pm 1/2}r_j \mapsto t^{\mp 1/2}\bar{r}_j $ 
\end{proposition}

\begin{proof}
Consider first the specialisation of (\ref{ops_OTeq:a}) to our weight at hand
at the singular point $ z=-1 $, and we have
\begin{multline} \label{VI_Bilinear:a}
   \Big\{ {l_N \over \kappa_N} - Nt^{-1} 
          - (N+1+\mu+\bar{\omega}){\kappa^2_{N-1} \over \kappa^2_N}{r_{N+1} \over r_N} 
          + \omega_1(1-t^{-1}) \Big\}^2
   \\
   + {\kappa^2_{N-1} \over \kappa^2_N}
     \left[ N+\mu+\bar{\omega} + {(N-1+\mu+\omega)\over t}{r_{N-1} \over r_N}
     \right]
   \\ \times
     \left[ {(N+\mu+\omega)\over t} + (N+1+\mu+\bar{\omega}){r_{N+1} \over r_N}
     \right]
   = \omega^2_1 \left({t-1 \over t}\right)^2 ,
\end{multline}
by using (\ref{VI_theta:a},\ref{VI_omega:a}). Similarly (\ref{ops_OTeq:a}) 
evaluated at $ z=-1/t $ yields
\begin{multline} \label{VI_Bilinear:b}
   \Big\{ {l_N \over \kappa_N} - N 
          - (N+1+\mu+\bar{\omega}){\kappa^2_{N-1} \over \kappa^2_N}{r_{N+1} \over r_N} 
          + \mu(t^{-1}-1) \Big\}^2
   \\
   + {\kappa^2_{N-1} \over \kappa^2_N}t^{-1}
     \left[ N+\mu+\bar{\omega} + (N-1+\mu+\omega){r_{N-1} \over r_N}
     \right]
   \\ \times
     \left[ N+\mu+\omega + (N+1+\mu+\bar{\omega}){r_{N+1} \over r_N}
     \right]
   = \mu^2 \left({t-1 \over t}\right)^2 .
\end{multline}
The first relation follow by eliminating $ l_N $ between (\ref{VI_Bilinear:a}) and
(\ref{VI_Bilinear:b}), whereas the second follows from an identical analysis to that
employed in the proof of Proposition \ref{propVI_dPa} but starting with the bilinear 
identity (\ref{ops_OTeq:b}).
\end{proof}

\begin{proposition}
The reflection coefficients also satisfy a system of $ 1/1 $ order recurrence relations
the first of which is
\begin{multline}\label{VI_1+1rRecur:a}
  \Big\{ -(N+\mu+\omega)(1-r_N\bar{r}_N)t\left[
  (N+1+\mu+\bar{\omega})r_{N+1}\bar{r}_N + (N-1+\mu+\bar{\omega})r_N\bar{r}_{N-1} \right]
  \\ + 2(N+\mu+\omega)^2r^2_N\bar{r}^2_N - (N+\mu+\omega)^2(t+1)r_N\bar{r}_N
     - 2(N+\mu+\omega)\bar{\omega}(t-1)r_N\bar{r}_N
  \\
     + (\mu-\bar{\omega})(\mu+\bar{\omega})(t-1) \Big\}
  \\ \times
  \\
  \Big\{ -(N+\mu+\omega)(1-r_N\bar{r}_N)t\left[
  (N+1+\mu+\bar{\omega})r_{N+1}\bar{r}_N + (N-1+\mu+\bar{\omega})r_N\bar{r}_{N-1} \right]
  \\ + 2(N+\mu+\omega)^2r^2_N\bar{r}^2_N - (N+\mu+\omega)^2(t+1)r_N\bar{r}_N
     + 2(N+\mu+\omega)\omega(t-1)r_N\bar{r}_N
  \\
     + (\mu-\omega)(\mu+\omega)(t-1) \Big\}
  \\
  = -\left[ 2(N+\mu+\omega)r_N\bar{r}_N+\bar{\omega}-\omega \right]^2(1-r_N\bar{r}_N)
  \\ \times
  \left[ (N+1+\mu+\bar{\omega})tr_{N+1}+(N+\mu+\omega)r_{N} \right]
  \\ \times
  \left[ (N+\mu+\omega)\bar{r}_{N}+(N-1+\mu+\bar{\omega})t\bar{r}_{N-1} \right] ,
\end{multline}
and the second is obtained from (\ref{VI_1+1rRecur:a}) with the replacements
$ \omega \leftrightarrow \bar{\omega} $ and
$ t^{\pm 1/2}r_j \leftrightarrow t^{\mp 1/2}\bar{r}_j $.    
\end{proposition}

\begin{proof}
The specialisation of (\ref{ops_OTeq:c}) to the weight (\ref{VI_wgt}) evaluated
at the singular point $ z=-1 $ is
\begin{multline} \label{VI_Bilinear:c}
   \Big\{ {l_N \over \kappa_N} - Nt^{-1} 
          + (N+\mu+\omega)t^{-1}{\kappa^2_{N-1} \over \kappa^2_N}
          + \omega_1(1-t^{-1}) \Big\}^2
   \\
   + {\kappa^2_{N-1} \over \kappa^2_N}
     \left[ (N+1+\mu+\bar{\omega})r_{N+1} + (N+\mu+\omega)t^{-1} r_N
     \right]
   \\ \times
     \left[ (N-1+\mu+\bar{\omega})\bar{r}_{N-1} + (N+\mu+\omega)t^{-1}\bar{r}_N
     \right]
   = \omega^2_1 \left({t-1 \over t}\right)^2 ,
\end{multline}
by using (\ref{VI_theta:a},\ref{VI_omega:a}). Similarly (\ref{ops_OTeq:c}) 
evaluated at $ z=-1/t $ yields
\begin{multline} \label{VI_Bilinear:d}
   \Big\{ {l_N \over \kappa_N} - N 
          + (N+\mu+\omega){\kappa^2_{N-1} \over \kappa^2_N}
          + \mu(t^{-1}-1) \Big\}^2
   \\
   + {\kappa^2_{N-1} \over \kappa^2_N}
     \left[ (N+1+\mu+\bar{\omega})r_{N+1} + (N+\mu+\omega)r_N
     \right]
   \\ \times
     \left[ (N-1+\mu+\bar{\omega})\bar{r}_{N-1} + (N+\mu+\omega)\bar{r}_N
     \right]
   = \mu^2 \left({t-1 \over t}\right)^2 .
\end{multline}
Again eliminating $ l_N $ between these two equations yields the recurrence relation
(\ref{VI_1+1rRecur:a}). The second follows in the same way starting with (\ref{ops_OTeq:d}).
\end{proof}

\begin{proposition}
The reflection coefficients satisfy an alternative system of $ 1/1 $ order recurrence 
relations the first of which is
\begin{multline}\label{VI_1+1rRecur:b}
  \Big[ (N+1+\mu+\bar{\omega})(N+\mu+\bar{\omega})tr_{N+1}\bar{r}_N 
  \\  - (N+1+\mu+\omega)(N+\mu+\omega)\bar{r}_{N+1}r_N
      + (\bar{\omega}-\mu)(\bar{\omega}+\mu)(t-1) \Big]
  \\ \times
  \Big[ (N+1+\mu+\bar{\omega})(N+\mu+\bar{\omega})tr_{N+1}\bar{r}_N 
  \\  - (N+1+\mu+\omega)(N+\mu+\omega)\bar{r}_{N+1}r_N
      + (\omega-\mu)(\omega+\mu)(t-1) \Big]
  \\ 
  = (\bar{\omega}-\omega)^2
  \left[ (N+1+\mu+\bar{\omega})tr_{N+1}+(N+\mu+\omega)r_{N} \right]
  \\ \times
  \left[ (N+1+\mu+\omega)\bar{r}_{N+1}+(N+\mu+\bar{\omega})t\bar{r}_{N} \right] ,
\end{multline}
and the second is again obtained from (\ref{VI_1+1rRecur:b}) with the replacements
$ \omega \leftrightarrow \bar{\omega} $ and
$ t^{\pm 1/2}r_j \leftrightarrow t^{\mp 1/2}\bar{r}_j $.    
\end{proposition}

\begin{proof}
The specialisation of (\ref{ops_OTeq:e}) to the weight (\ref{VI_wgt}) evaluated
at the singular point $ z=-1 $ is
\begin{multline} \label{VI_Bilinear:e}
   \Big\{ {\bar{l}_{N+1} \over \kappa_{N+1}}
          + (N+\mu+\omega)\bar{r}_{N+1}r_N
          + \omega_1+(\mu-i\omega_2)t \Big\}^2
   \\
   = \left[ (N+1+\mu+\bar{\omega})tr_{N+1} + (N+\mu+\omega)r_N \right]
   \\ \times
     \left[ (N+1+\mu+\omega)\bar{r}_{N+1} + (N+\mu+\bar{\omega})t\bar{r}_N \right]
     + \omega^2_1 (t-1)^2 ,
\end{multline}
by using (\ref{VI_theta:a},\ref{VI_omega:a}). Similarly (\ref{ops_OTeq:e}) 
evaluated at $ z=-1/t $ yields
\begin{multline} \label{VI_Bilinear:f}
   \Big\{ {\bar{l}_{N+1} \over \kappa_{N+1}}
          + (N+\mu+\omega)\bar{r}_{N+1}r_N
          + \bar{\omega}+\mu t \Big\}^2
   \\
   = t\left[ (N+1+\mu+\bar{\omega})r_{N+1} + (N+\mu+\omega)r_N \right]
   \\ \times
     \left[ (N+1+\mu+\omega)\bar{r}_{N+1} + (N+\mu+\bar{\omega})\bar{r}_N \right]
     + \mu^2 (t-1)^2 ,
\end{multline}
Again eliminating $ \bar{l}_{N+1} $ between these two equations yields the recurrence relation
(\ref{VI_1+1rRecur:b}). The second follows in the same way starting with (\ref{ops_OTeq:f}).
\end{proof}

\begin{remark}
Note that the recurrence system (\ref{VI_2+0rRecur:a}) and its partner is quadratic
in $ r_{N+1} $, $ r_{N-1} $ and $ \bar{r}_{N+1} $, $\bar{r}_{N-1} $, the system
(\ref{VI_1+1rRecur:a}) and its partner is also quadratic in $ r_{N+1}, \bar{r}_{N-1} $
and $ \bar{r}_{N+1} $, $ r_{N-1} $, and likewise (\ref{VI_1+1rRecur:b}) is quadratic in 
$ r_{N+1} $, $ \bar{r}_{N+1} $. This renders them less useful in practical iterations
than the higher order systems that are linear in the highest difference. By raising the 
order of one of the variables by one we can obtain a recurrence linear in the highest
difference.
\end{remark}

\begin{corollary}
The reflection coefficients satisfy a system of a $ 2/1 $ order recurrence relation
\begin{multline}\label{VI_2+1rRecur:a}
  (N+1+\mu+\bar{\omega})(\bar{\omega}-\omega)t(1-r_N\bar{r}_N)r_{N+1}
 \\
 + (N-1+\mu+\omega)[2(N+\mu+\omega)r_N\bar{r}_N+\bar{\omega}-\omega]r_{N-1}
 \\
 - (N-1+\mu+\bar{\omega})(2N+2\mu+2\omega_1)tr^2_N\bar{r}_{N-1}
 \\
 + \big[ (\bar{\omega}-\omega)N(t+1)-(2\mu+2\omega_1)[\mu(1-t)+\omega t-\bar{\omega}]
   \big]r_N = 0 ,
\end{multline}
and a $ 1/2 $ order recurrence relation which is again obtained from 
(\ref{VI_2+1rRecur:a}) with the replacements $ \omega \leftrightarrow \bar{\omega} $ and
$ t^{\pm 1/2}r_j \leftrightarrow t^{\mp 1/2}\bar{r}_j $.
\end{corollary}

\begin{proof}
The solutions for the sub-leading coefficient $ l_N, \bar{l}_N $ that arise from the 
simultaneous solution of (\ref{VI_Bilinear:c},\ref{VI_Bilinear:d}) and 
(\ref{VI_Bilinear:e},\ref{VI_Bilinear:f}) respectively are given by
\begin{multline}
  t{l_N \over \kappa_N} 
  = \Big\{ (N+\mu+\omega)t(1-r_N\bar{r}_N)
  \\ \times
      \left[ (N+1+\mu+\bar{\omega})r_{N+1}\bar{r}_N+(N-1+\mu+\bar{\omega})r_N\bar{r}_{N-1}
      \right]
  \\
    +(N+\mu+\omega)[N(t+1)-\mu(1-t)-\omega t+\bar{\omega}]r_N\bar{r}_N
    +(\omega+\mu)[\mu(1-t)+\omega t-\bar{\omega}] \Big\}
  \\
    \div [2(N+\mu+\omega)r_N\bar{r}_N+\bar{\omega}-\omega] ,
  \label{VI_lSoln:c}
\end{multline}
\begin{multline}
  t{l_N \over \kappa_N}
  = \Big\{ (N+\mu+\omega)
      \left[ (N-1+\mu+\bar{\omega})tr_{N}\bar{r}_{N-1}-(N-1+\mu+\omega)\bar{r}_{N}r_{N-1}
      \right]
  \\
    + (\omega+\mu)[\mu(1-t)+\omega t-\bar{\omega}] \Big\}\Big/(\bar{\omega}-\omega),
    \quad \text{if $ \bar{\omega} \neq \omega $} ,
  \label{VI_lSoln:d} 
\end{multline}
and the corresponding expression for $ \bar{l}_N/\kappa_N $ under the above replacements.
Equating these two forms then leads to (\ref{VI_2+1rRecur:a}).
\end{proof}

The systems of recurrences that we have found are in fact equivalent to the 
discrete Painlev\'e equation associated with the 
degeneration of the rational surface $ D^{(1)}_4 \to D^{(1)}_5 $ and we give
our first demonstration of this fact here.
\begin{proposition}\label{propVI_dPa}
The $ N $-recurrence for the reflection coefficients of the orthogonal 
polynomial system with the weight (\ref{VI_wgt}) is governed by either of two 
systems of coupled first order discrete Painlev\'e equations (\ref{dPV:a}), (\ref{dPV:b}).
This first is
\begin{align}                                                                 
  g_{N+1}g_N                                                                
  & = t{(f_N+N)(f_N+N+2\mu) \over f_N(f_N-2\omega_1)} ,
  \label{VI_gRecur} \\       
  f_N+f_{N-1}      
  & = 2\omega_1+{N-1+\mu+\omega \over g_N-1}
      +{(N+\mu+\bar{\omega})t \over g_N-t},
  \label{VI_fRecur}
\end{align}
subject to the initial conditions
\begin{equation}
  g_1 = t{ \mu+\omega +(1+\mu+\bar{\omega})r_1 \over 
           \mu+\omega +(1+\mu+\bar{\omega})tr_1 } ,
  \quad f_0 = 0 .
\end{equation}
The transformations relating these variables to the reflection coefficients are
given by
\begin{align}
   g_N & =
  t{ N-1+\mu+\omega +(N+\mu+\bar{\omega})
                     \dfrac{r_N}{r_{N-1}}
     \over 
     N-1+\mu+\omega +(N+\mu+\bar{\omega})t
                     \dfrac{r_N}{r_{N-1}} } ,
  \label{VI_gXfm} \\
   f_N & = 
  {1 \over 1-t}\left[ 
   t{l_N \over \kappa_N} - N -(N+1+\mu+\bar{\omega})(1-r_N\bar{r}_N)t{r_{N+1} \over r_{N}}
               \right] .
  \label{VI_fXfm}
\end{align}
The second system is
\begin{align}                                                                 
  \bar{g}_{N+1}\bar{g}_N                                                                
  & = t^{-1}{(\bar{f}_N+N)(\bar{f}_N+N+2\omega_1) \over \bar{f}_N(\bar{f}_N-2\mu)} ,
  \label{VI_gRecurC} \\       
  \bar{f}_N+\bar{f}_{N-1}      
  & = 2\mu+{N+\mu+\omega \over \bar{g}_N-1}
      +{(N-1+\mu+\bar{\omega})t^{-1} \over \bar{g}_N-t^{-1}},
  \label{VI_fRecurC}
\end{align}
subject to the initial conditions
\begin{equation}
  \bar{g}_1 = {\mu+\bar{\omega} +(1+\mu+\omega)t^{-1}\bar{r}_1 \over 
           \mu+\bar{\omega} +(1+\mu+\omega)\bar{r}_1 } ,
  \quad \bar{f}_0 = 0 .
\end{equation}
The transformations relating these variables to the reflection coefficients are
given by
\begin{align}
   \bar{g}_N & =
   { N-1+\mu+\bar{\omega} +(N+\mu+\omega)t^{-1}
                     \dfrac{\bar{r}_N}{\bar{r}_{N-1}}
     \over 
     N-1+\mu+\bar{\omega} +(N+\mu+\omega)
                     \dfrac{\bar{r}_N}{\bar{r}_{N-1}} } ,
  \label{VI_gXfmC} \\
   \bar{f}_N & = 
  {1 \over 1-t}\left[ 
   -t{l_N \over \kappa_N}+Nt 
      +(N-1+\mu+\bar{\omega})(1-r_N\bar{r}_N)t{\bar{r}_{N-1} \over \bar{r}_{N}}
               \right] .
  \label{VI_fXfmC}
\end{align} 
\end{proposition}

\begin{proof}
Consolidating each of (\ref{VI_Bilinear:a}) and (\ref{VI_Bilinear:b}) into two 
terms and taking their ratio then leads to (\ref{VI_gRecur}) after
utilising the definitions (\ref{VI_gXfm},\ref{VI_fXfm}). The second member of the
recurrence system (\ref{VI_fRecur}) follows from the relation
\begin{multline}
  {l_{N+1} \over \kappa_{N+1}} + {l_N \over \kappa_N}
  \\
  = (N+2+\mu+\bar{\omega})(1-r_{N+1}\bar{r}_{N+1}){r_{N+2} \over r_{N+1}}
  + (N+\mu+\omega)t^{-1}{r_{N} \over r_{N+1}} 
  - (N+1+\mu+\bar{\omega})r_{N+1}\bar{r}_N
  \\
  - 2\omega_1 - 2\mu t^{-1} + (N+1+\mu+\bar{\omega})(1+t^{-1}) ,
\end{multline}
which results from a combination of (\ref{VI_lSoln:a}) and (\ref{ops_l}),
and the definition (\ref{VI_fXfm}). All the results for the second system follow
by applying identical reasoning starting with (\ref{ops_OTeq:b}).
\end{proof}

\subsection{Evaluations in terms of generalised hypergeometric functions}

In the special case $ \xi=0 $ of the $ U(N) $ average (\ref{ops_Uavge}) with weight
(\ref{VI_wgt}), we know from \cite{FW_2002b} that an evaluation in terms of a 
generalised hypergeometric function $ {}^{\vphantom{(1)}}_{2}F^{(1)}_{1} $
is possible. Let us first recall the definition of the latter. Given a partition
$ \kappa = (\kappa_1,\kappa_2, \ldots, \kappa_N) $ such that 
$ \kappa_1 \geq \kappa_2 \geq \cdots \geq \kappa_N \geq 0 $ one defines the
generalised, multi-variable hypergeometric function through a series representation
\cite{Ya_1992,Ka_1993}
\begin{equation}
  {}^{\vphantom{(1)}}_{p}F^{(1)}_{q}(a_1,\ldots,a_p;b_1,\ldots,b_q;t_1,\ldots,t_N)
  = \sum^{\infty}_{\kappa \geq 0} 
    {[a_1]^{(1)}_{\kappa}\cdots [a_p]^{(1)}_{\kappa} \over 
     [b_1]^{(1)}_{\kappa}\cdots [b_q]^{(1)}_{\kappa}}
    {s_{\kappa}(t_1,\ldots,t_N) \over h_{\kappa}} ,
\label{pFq_defn}
\end{equation}
for $ p,q \in \ZZ_{\geq 0} $. Here the generalised Pochhammer symbols are
\begin{equation}
   [a]^{(1)}_{\kappa} := \prod^{N}_{j=1}(a-j+1)_{\kappa_j} ,
\end{equation}
the hook length is
\begin{equation}
   h_{\kappa} = \prod_{(i,j) \in \kappa} [a(i,j)+l(i,j)+1] ,
\end{equation}
where $ a(i,j), l(i,j) $ are the arm and leg lengths of the $ (i,j) $th
box in the Young diagram of the partition $ \kappa $, and
$ s_{\kappa}(t_1,\ldots,t_N) $ is the Schur symmetric polynomial of 
$ N $ variables. The superscript $ (1) $ distinguishes these functions from the 
single variable $ N=1 $ functions and also indicates that they are a special 
case of a more general function parameterised by an arbitrary complex number 
$ d \neq 1 $.

With this definition recalled, the result of \cite{FW_2002b} (see also 
\cite{Fo_1992}) reads
\begin{multline}
   \Big\langle \prod^{N}_{l=1}
   z_l^{-\mu-\omega} (1+z_l)^{2\omega_1}(1+tz_l)^{2\mu} \Big\rangle_{U(N)} \\
  = \prod^{N-1}_{j=0}{j!\Gamma(2\omega_1+j+1) \over 
                      \Gamma(1+\mu+\omega+j)\Gamma(1-\mu+\bar{\omega}+j)} \\
  \times
    {}^{\vphantom{(1)}}_{2}F^{(1)}_{1}(-2\mu,-\mu-\omega;N-\mu+\bar{\omega};t_1,\ldots,t_N)
       |_{t_1=\ldots =t_N=t} ,
\label{VI_2F1}
\end{multline}
subject to $ \Re(\omega_1) > -\half $ and $ |t| < 1 $.
The new observation we make here is that the reflection coefficients determining 
(\ref{VI_2F1}) can similarly be written in terms of the 
$ {}^{\vphantom{(1)}}_{2}F^{(1)}_{1} $ function.
\begin{proposition}
With $ I_N[w] $ in (\ref{ops_I0}) given by (\ref{VI_2F1}), the corresponding
reflection coefficients are given by
\begin{multline}
   r_{N}
    = (-1)^{N}{(\mu+\omega)_{N} \over (1-\mu+\bar{\omega})_{N}}
   \\ \times
 {{}^{\vphantom{(1)}}_{2}F^{(1)}_{1}(-2\mu,1-\mu-\omega;N+1-\mu+\bar{\omega};t_1,\ldots,t_N)
      \over 
  {}^{\vphantom{(1)}}_{2}F^{(1)}_{1}(-2\mu,-\mu-\omega;N-\mu+\bar{\omega};t_1,\ldots,t_N) 
  }\Bigg|_{t_1=\ldots =t_N=t},
  \label{VI_genH:a}
\end{multline}
\begin{multline}
   \bar{r}_{N}
    = (-1)^{N}{(-\mu+\bar{\omega})_{N} \over (1+\mu+\omega)_{N}}
   \\ \times
 {{}^{\vphantom{(1)}}_{2}F^{(1)}_{1}(-2\mu,-1-\mu-\omega;N-1-\mu+\bar{\omega};t_1,\ldots,t_N)
      \over 
  {}^{\vphantom{(1)}}_{2}F^{(1)}_{1}(-2\mu,-\mu-\omega;N-\mu+\bar{\omega};t_1,\ldots,t_N)
  }\Bigg|_{t_1=\ldots =t_N=t}.
  \label{VI_genH:b}
\end{multline}
\end{proposition}
\begin{proof}
For $ \epsilon=0,\pm 1 $ we define the Toeplitz determinants or $ U(N) $ averages
\begin{equation}
  I^{\epsilon}_n[w] 
  := \det\left[ \int_{\TT}\frac{d\zeta}{2\pi i\zeta}w(\zeta)\zeta^{\epsilon-j+k}
         \right]_{0\leq j,k\leq n-1}
   = \Big\langle \prod^{n}_{l=1}z^{\epsilon}_lw(z_l) \Big\rangle_{U(n)} . 
\end{equation} 
From the Szeg\"o theory we know
\begin{equation}
   r_n = (-1)^n\frac{I^{1}_n[w]}{I^{0}_n[w]}, \quad
   \bar{r}_n = (-1)^n\frac{I^{-1}_n[w]}{I^{0}_n[w]} .
\end{equation}
The result now follows from (\ref{VI_2F1}).
\end{proof}

According to the definition (\ref{pFq_defn}), 
$ {}^{\vphantom{(1)}}_{2}F^{(1)}_{1} $ is normalised so that 
\begin{equation}
   {}^{\vphantom{(1)}}_{2}F^{(1)}_{1}(a,b;c;t_1,\ldots,t_N)|_{t_1=\ldots =t_N=0} = 1 .
\label{2F1_norm}
\end{equation}
At the special point when each $ t_i $ equals unity, the analog of the Gauss summation 
gives the gamma function evaluation \cite{Ya_1992}
\begin{multline}
  {}^{\vphantom{(1)}}_{2}F^{(1)}_{1}(-2\mu,-\mu-\omega;N-\mu+\bar{\omega};t_1,\ldots,t_N)
       |_{t_1=\ldots =t_N=1}
  \\ =
    \prod^{N}_{j=1}{\Gamma(j+2\mu+2\omega_1)\Gamma(j-\mu+\bar{\omega}) \over  
                      \Gamma(j+2\omega_1)\Gamma(j+\mu+\bar{\omega})} ,
\label{2F1_Gsum}
\end{multline}
when $ \Re(\mu+\omega_1) > -\half $, $ \Re(-\mu+\bar{\omega}) > -1 $.
 
The result (\ref{2F1_norm}) tells us that at $ t=0 $
\begin{gather}
  r_{N} = (-1)^{N}{(\mu+\omega)_{N} \over (1-\mu+\bar{\omega})_{N}} ,\quad
  \bar{r}_{N} = (-1)^{N}{(-\mu+\bar{\omega})_{N} \over (1+\mu+\omega)_{N}} ,
  \label{VI_t=zero}
\end{gather}
while (\ref{2F1_Gsum}) tells us that at $ t=1 $
\begin{equation}
  r_{N} = (-1)^{N}{(\mu+\omega)_{N} \over (1+\mu+\bar{\omega})_{N}} ,\quad
  \bar{r}_{N} = (-1)^{N}{(\mu+\bar{\omega})_{N} \over (1+\mu+\omega)_{N}} .
\end{equation}
We note that the corresponding values of $ l_N $ can be calculated as
\begin{equation}
  \frac{l_{N}}{\kappa_N} = -{(\mu+\omega)N \over (N-\mu+\bar{\omega})}, \quad
  \frac{l_{N}}{\kappa_N} = -{(\mu+\omega)N \over (N+\mu+\bar{\omega})} .
\end{equation}

\subsection{Two remarks}

We conclude this section with two remarks. The first relates to the limit 
transition from the average (\ref{ops_Uavge}) with weight (\ref{VI_wgt}) to the
average
\begin{equation}
   \Big\langle \prod^N_{l=1}z^{(\mu-\nu)/2}_l|1+z_l|^{\mu+\nu}e^{tz_l} \Big\rangle_{U(N)} ,
\end{equation}
studied from the viewpoint of bi-orthogonal polynomials in \cite{FW_2003b}.
The latter can be obtained as a degeneration of the former by the replacements
$ \omega+\mu \mapsto \nu, \bar{\omega}-\mu \mapsto \mu, t \mapsto t/2\mu $ 
and then taking the limit $ \mu \to \infty $. The coefficients of the orthogonal 
polynomials $ r_N, l_N $ remain of $ {\rm O}(1) $ in this limit. 
Then we see the explicit degeneration of the following equations - 
(\ref{VI_lRecur}) $\to$ Equation (4.23)\cite{FW_2003b},
the recurrence relations (\ref{VI_1+1rRecur:a}) $\to$ Equation (4.60)\cite{FW_2003b},
(\ref{VI_2+1rRecur:a}) $\to$ Equation (4.9)\cite{FW_2003b} and its conjugate to 
Equation (4.10)\cite{FW_2003b} modulo the identity Equation (4.5)\cite{FW_2003b},
and the hypergeometric functions (\ref{VI_2F1}) $\to$ Equation (4.24)\cite{FW_2003b},
(\ref{VI_genH:a}) $\to$ Equation (4.26)\cite{FW_2003b}, and
(\ref{VI_genH:b}) $\to$ Equation (4.27)\cite{FW_2003b}.

The second remark relates to the choices of parameters, noted below (\ref{VI_wgt}),
for which the weight is real and positive, and consequently 
$ \bar{r}_n=r_n $. Let us suppose furthermore that $ \omega_2=0 $, $ \xi=0 $.
Then with $ t=e^{i\phi} $, (\ref{VI_wgt}) reads
\begin{equation}
  w(e^{i\theta}) = |2\cos\shalf\theta|^{2\omega}|2\cos\shalf(\theta+\phi)|^{2\mu} .
\label{VI_realwgt}
\end{equation}
It follows from this that $ w_{-1}=tw_{1} $, and this in (\ref{VI_rInitial})
tells us that $ \bar{r}_1 = t r_ 1 $. This initial value, together with the
initial value given by the first equation in (\ref{VI_rInitial}), allows a 
structural formula for $ r_n $ in this case to be obtained.

\begin{corollary}\label{realRC}
Let (\ref{VI_wgt}) be specialised to (\ref{VI_realwgt}). Then the reflection 
coefficient $ r_n $, which is related to $ \bar{r}_n $ by complex conjugation,
has the form $ r_n = t^{-n/2}x_n $, where the $ x_n $ are real. 
\end{corollary}
\begin{proof}
Setting $ \omega_2=0 $ in (\ref{VI_2ndRR}) we note this can be rearranged as
\begin{equation}
   (n+1+\mu+\omega)\left[ t{r_{n+1} \over r_n} - {\bar{r}_{n+1} \over \bar{r}_n}
                   \right]
 + (n-1+\mu+\omega)\left[ {r_{n-1} \over r_n} - t{\bar{r}_{n-1} \over \bar{r}_n}
                   \right] = 0 .
\end{equation}
It is easy to verify that this has the solution $ \bar{r}_n=t^nr_n $ which is 
furthermore consistent with the initial conditions. The result now follows from 
the fact that $ \bar{r}_n $ is the complex conjugate of $ r_n $. 
\end{proof}

\section{The $\tau$-function Theory for \PVI}
\label{tausection}
\setcounter{equation}{0}

In a previous study \cite{FW_2003a} the average (\ref{ops_Uavge}) with weight 
(\ref{VI_wgt}) has been shown to satisfy recurrences involving \dPV, distinct
from those isolated in Proposition \ref{propVI_dPa}. Here we will present
the explicit transformation between the two systems. We will also show how the
relationship between the $ U(N) $ average and an average over the Jacobi unitary
ensemble, studied from the viewpoint of $ \tau$-function theory in \cite{FW_2002b},
can be used to deduce a further characterisation involving \dPV.

The study \cite{FW_2003a} is based on the Okamoto $ \tau$-function theory of
Painlev\'e systems. In the case of \PVI, a Hamiltonian $ H $ is defined
in terms of coordinate and momenta variables $ q,p $, a time variable $ t $, and
parameters $ \alpha_0,\alpha_1,\alpha_2,\alpha_3,\alpha_4\in \CC $ subject to the
constraint
\begin{equation}
    \alpha_0+\alpha_1+2\alpha_2+\alpha_3+\alpha_4 = 1 ,
\end{equation}
according to
\begin{multline}
  K := t(t-1)H \\
    =  q(q-1)(q-t)p^2 
     - \left[ \alpha_4(q-1)(q-t)+\alpha_3q(q-t)+(\alpha_0-1)q(q-1) \right]p
    \\
     + \alpha_2(\alpha_1+\alpha_2)(q-t) .
\label{VI_Ham}
\end{multline}
The time evolution of $ q $ and $ p $ is governed by the Hamilton equations
\begin{equation}\label{VI_Hdyn}
  {dq \over dt} =  {\partial H\over \partial p} , \qquad
  {dp \over dt} = -{\partial H\over \partial q} ,
\end{equation}
with the relationship to \PVI coming from the fact that eliminating $ p $ gives
the sixth Painlev\'e equation in $ q $. A crucial quantity in the development of
this viewpoint given in \cite{Ok_1987a} is the $\tau$-function, defined so that
\begin{equation}
    H = {d \over dt}\log \tau .
\end{equation}
It was shown in \cite{FW_2002b} that the $ U(N) $ average (\ref{ops_Uavge}) with
weight (\ref{VI_wgt}) is a  $\tau$-function, for the \PVI system with the 
parameters
\begin{equation}
  (\alpha_0,\alpha_1,\alpha_2,\alpha_3,\alpha_4) 
   = \Big( N+1+2\omega_1,N+2\mu,-N,-\mu-\omega,-\mu-\bar{\omega} \Big) ,
\label{VI_CyUEparam}
\end{equation}
and so we can write
\begin{multline}
  \tau^{\rm VI}[N](t;\mu,\omega_1,\omega_2;\xi) \\
 = \Big\langle \prod^N_{l=1}(1-\xi\chi^{(l)}_{(\pi-\phi,\pi)})e^{\omega_2\theta_l}
                        |1+z_l|^{2\omega_1}\left(\frac{1}{tz_l}\right)^{\mu}(1+tz_l)^{2\mu}
   \Big\rangle_{U(N)} .
\label{VI_Uint}
\end{multline}

The algebraic approach used in \cite{FW_2003a} makes use of a particular shift 
operator, or Schlesinger transformation $ L $, constructed from compositions of 
fundamental reflection operators and Dynkin diagram automorphisms of the 
\PVI affine Weyl symmetry group 
$ W_a(D^{(1)}_4) = \langle s_0,s_1,s_2,s_3,s_4,r_1,r_3,r_4 \rangle $.
Application of the operator $ L $ allows for $ \tau^{\rm VI}[N] $ to be computed
by a recurrence scheme in $ N $ involving auxiliary quantities which satisfy the 
\dPV recurrence.

Explicitly, the operator with the property of incrementing $ N $ while leaving the 
other parameters unchanged is $ L^{-1}_{01} = r_1s_0s_1s_2s_3s_4s_2 $. On the 
root system parameters $ \alpha_j $ in (\ref{VI_Ham}) it has the action
\begin{equation}
  L^{-1}_{01}: 
  \alpha_0 \mapsto \alpha_0+1, \alpha_1 \mapsto \alpha_1+1, \alpha_2 \mapsto \alpha_2-1 ,
\end{equation}
while $ \alpha_3 $ and $ \alpha_4 $ remain unchanged. Studying the action of 
$ L^{-1}_{01} $ on $ \tau^{\rm VI}[N] $ in the context of the Okamoto theory, the
following result was obtained in \cite{FW_2003a}.
\begin{proposition}[\cite{FW_2003a}]\label{VI_L01dPV}
Let $\{g_N,f_N\}_{N=0,1,\dots}$, satisfy the discrete Painlev\'e coupled difference 
equations associated with the degeneration of the rational surface 
$ D^{(1)}_4 \to D^{(1)}_5 $
\begin{align}
  g_{N+1}g_N 
  & = {t\over t-1}{(f_N+N+1)(f_N+N+1+\mu+\bar{\omega}) \over f_N(f_N-\mu-\omega)} ,
  \label{VI_L01Recur:a} \\
  f_N+f_{N-1} 
  & = \mu+\omega+{N+2\mu \over g_N-1}
      +{(N+1+2\omega_1)t \over t(g_N-1)-g_N},
  \label{VI_L01Recur:b}
\end{align}
where $ t=1/(1-e^{i\phi}) $ subject to the initial conditions
\begin{equation*}
  g_0 = {q_0 \over q_0-1}, \quad
  f_0 = (1+\mu+\bar{\omega})(q_0-1)+(\mu+\omega)q_0-(2\omega_1+1){q_0(q_0-1) \over q_0-t} ,
\end{equation*}
with
\begin{equation}
  q_0 = \frac{1}{2}\left( 1+\frac{i}{\mu}{d \over d\phi}\log e^{i\mu\phi}T_1(e^{i\phi})
                   \right) .
\label{VI_L01Tinitial}
\end{equation}
Define $\{q_N,p_N\}_{N=0,1,\dots}$ by
\begin{equation*}
  q_N = {g_N \over g_N-1},
\end{equation*}
\begin{multline}
  p_N = {(g_N-1)^2 \over g_N}f_N \\
        -(N+1+\mu+\bar{\omega}){g_N-1 \over g_N}-(\mu+\omega)(g_N-1)
                 +(N+1+2\omega_1){g_N-1 \over t+(1-t)g_N} .
\nonumber 
\end{multline}
Then with $T_0(e^{i\phi}) = 1$ and $T_1(e^{i\phi})=w_0(e^{i\phi}) $ as given by
(\ref{VI_toepM},\ref{VI_toepM2}), $\{T_N\}_{N=2,3,\dots}$ is specified by 
the recurrence
\begin{multline}
  -(N+\mu+\omega)(N+\mu+\bar{\omega})
  {T_{N+1}T_{N-1} \over T_N^2 } \\
  = q_N(q_N-1)p^2_N+(2\mu+2\omega_1)q_Np_N-(\mu+\bar{\omega})p_N-N(N+2\mu+2\omega_1) .
\label{VI_L01Trecur}
\end{multline}
\end{proposition}

An immediate question is the relationship between the Hamiltonian variables 
$ q_N, p_N $ in Proposition \ref{VI_L01dPV}, and the reflection coefficients
$ r_N, \bar{r}_N $ relating to $ \tau^{\rm VI}[N] $ as studied in Section 3.
In fact the quantities are linked by systems of equations given in the 
following result.
\begin{proposition}
The transformations linking the Hamiltonian variables $ q_N, p_N $ in Proposition
\ref{VI_L01dPV} to the reflection coefficients $ r_N, \bar{r}_N $ for the system 
of orthogonal polynomials with the weight (\ref{VI_wgt}) are given implicitly by
\begin{gather}
   q_Np_N+\mu+\bar{\omega}
   = {(N+\mu+\bar{\omega})r_N\bar{r}_N \over (N+\mu+\bar{\omega})r_N\bar{r}_N-\mu+\omega}
     {1\over q_N-1} \nonumber \\
  \times
     \left[ (N+2\omega_1)(q_N-1)
            - t{l_N \over \kappa_N}+Nt
   +(N+1+\mu+\bar{\omega})(1-r_N\bar{r}_N)t{r_{N+1} \over r_N} \right] ,
   \label{VIH_Ops:a} \\
   = (N+\mu+\bar{\omega})[ (N+\mu+\omega)r_N\bar{r}_N-\mu+\bar{\omega} ] \nonumber \\
  \times
     { q_N \over (N+2\omega_1)q_N
     +t\dfrac{l_N}{\kappa_N}-Nt
   -(N-1+\mu+\bar{\omega})(1-r_N\bar{r}_N)t
    \dfrac{\bar{r}_{N-1}}{\bar{r}_N}} ,
  \label{VIH_Ops:b} \\
  (q_N-1)p_N+\mu+\omega
   = (N+\mu+\omega)[ (N+\mu+\bar{\omega})r_N\bar{r}_N-\mu+\omega ] \nonumber \\
  \times
     { q_N-1 \over (N+2\omega_1)(q_N-1)
           - t\dfrac{l_N}{\kappa_N}+Nt
   +(N+1+\mu+\bar{\omega})(1-r_N\bar{r}_N)t\dfrac{r_{N+1}}{r_N} } ,
  \label{VIH_Ops:c} \\
   ={(N+\mu+\omega)r_N\bar{r}_N \over (N+\mu+\omega)r_N\bar{r}_N-\mu+\bar{\omega}}
     {1\over q_N} \nonumber \\
  \times
     \left[ (N+2\omega_1)q_N
    +t{l_N \over \kappa_N}-Nt
    -(N-1+\mu+\bar{\omega})(1-r_N\bar{r}_N)t{\bar{r}_{N-1} \over \bar{r}_N} \right] .
  \label{VIH_Ops:d}
\end{gather}
\end{proposition}
\begin{proof}
We require in addition to the primary shift operator
$ L^{-1}_{01} $ generating the $ N \mapsto N+1 $ sequence another operator which 
has the action $ i\omega_2 \mapsto i\omega_2-1 $. This is
the secondary shift operator $ T^{-1}_{34} = r_1s_4s_2s_0s_1s_2s_4 $ and has the
action $ T^{-1}_{34}: \alpha_3 \to \alpha_3+1, \alpha_4 \to \alpha_4-1 $.
From Table 1 of \cite{FW_2002b} we compute the actions of $ T^{-1}_{34}, T_{34} $ on
the Hamiltonian to be
\begin{multline*}
  T^{-1}_{34}\cdot K_n - K_n 
  = -q_n(q_n-1)p_n \\
    +(\alpha_0+\alpha_4-1)(q_n-1)
    -(\alpha_2+\alpha_3)(\alpha_1+\alpha_2+\alpha_3){q_n-1 \over (q_n-1)p_n-\alpha_3},
\end{multline*}
\begin{multline*}
  T_{34}\cdot K_n - K_n 
  = -q_n(q_n-1)p_n \\
    +(\alpha_0+\alpha_3-1)q_n
    -(\alpha_2+\alpha_4)(\alpha_1+\alpha_2+\alpha_4){q_n \over q_np_n-\alpha_4}.
\end{multline*}
However
\begin{equation*}
  T^{-1}_{34}\cdot K_n - K_n = t(t-1){d \over dt}\log{I^{1}_n \over I^{0}_n}
  = t(t-1){d \over dt}\log r_n, 
\end{equation*}
and we employ the results of \ref{ops_rdot} and the evaluation of the
coefficient functions in (\ref{VI_omega:a},\ref{VI_omega:b}) to arrive at
\begin{align*}
   (t-1){\dot{r}_N \over r_N} & = {l_N \over \kappa_N}-N
   -(N+1+\mu+\bar{\omega})(1-r_N\bar{r}_N){r_{N+1} \over r_N} ,
   \\
   (t-1){\dot{\bar{r}}_N \over \bar{r}_N} & = -{l_N \over \kappa_N}+N
   +(N-1+\mu+\bar{\omega})(1-r_N\bar{r}_N){\bar{r}_{N-1} \over \bar{r}_N} .
\end{align*}
In addition we note that after recalling (\ref{ops_I0}), (\ref{VI_L01Trecur})
factorises into
\begin{equation*}  
   (N+\mu+\omega)(N+\mu+\bar{\omega})r_N\bar{r}_N 
  = [q_Np_N+\mu+\bar{\omega}][(q_N-1)p_N+\mu+\omega] .
\end{equation*}  
The stated results, (\ref{VIH_Ops:a}-\ref{VIH_Ops:d}), then follow.
\end{proof}

There is another perspective on $ \tau^{\rm VI}[N] $ for which the Okamoto
$\tau$-function theory can be used to provide a recurrence system based
on \dPV distinct from that in Proposition \ref{VI_L01dPV}. The starting point,
used extensively in \cite{FW_2002b}, is to obtain $ \tau^{\rm VI}[N]|_{\xi=0} $
as specified by \ref{VI_Uint} via the projection $ (-1,1) \to \TT $ of an
average over the Jacobi unitary ensemble. This relates (\ref{VI_Uint}) to
the \PVI system with the parameters
\begin{equation}
  (\alpha_0,\alpha_1,\alpha_2,\alpha_3,\alpha_4) 
   = \Big( 1-\mu-\omega,N+2\mu,-N, -\mu-\bar{\omega},N+2\omega_1 \Big) .
\label{VI_JUEparam}
\end{equation} 
Sequences of the Hamiltonian variables $ \{ q_n,p_n,H_n,\tau_n \}_{n=0,1,\ldots} $
are now generated by the shift operator 
$ L^{-1}_{14} = r_3s_1s_4s_2s_0s_3s_2 $. It has the action
$ L^{-1}_{14}: 
  \alpha_1 \mapsto \alpha_1+1, \alpha_2 \mapsto \alpha_2-1, \alpha_4 \mapsto \alpha_4+1 $.
Using the methods of \cite{FW_2002b} we have the following result.
\begin{lemma}\label{VI_L14seq}
The sequence of auxiliary variables $ \{g_n,f_n\}_{n=0,1,\ldots} $ defined by
\begin{equation}
  g_n := {q_n-t \over q_n - 1},
  \label{VI_L14gDefn}
\end{equation}
\begin{multline}
  f_n := {1 \over 1-t}\bigg[ (q_n-t)(q_n-1)p_n 
  \label{VI_L14fDefn} \\
  \phantom{{1 \over 1-t}\bigg[} 
       + (1-\alpha_0-\alpha_2)(q_n-1)-\alpha_3(q_n-t)
                    - \alpha_4{(q_n-t)(q_n-1) \over q_n} \bigg],
\end{multline}
generated by the shift operator $ L^{-1}_{14} $ satisfies the discrete Painlev\'e 
equations (\ref{dPV:a}), (\ref{dPV:b}).
\end{lemma}
\begin{proof}
Using the action of the fundamental reflections and Dynkin diagram automorphisms
given in Table 1 of \cite{FW_2002b} we compute the action of $ L^{-1}_{14} $ on
$ q $ and write it in the following way,
\begin{multline*}
{(q-t)(\hat{q}-t) \over (q-1)(\hat{q}-1)} = 
 t[ q(q-1)(q-t)p+(\alpha_1+\alpha_2)q^2
                -((\alpha_0+\alpha_1+\alpha_2)t-\alpha_0-\alpha_4)q-\alpha_4t ] \\
 \times
  [ q(q-1)(q-t)p+(\alpha_1+\alpha_2)q^2
                -((\alpha_1+\alpha_2)t-\alpha_4)q-\alpha_4t ] \\
 \div
  [ q(q-1)(q-t)p+(\alpha_1+\alpha_2)q^2
                -(-\alpha_4t+\alpha_1+\alpha_2)q-\alpha_4t ] \\
 \div
  [ q(q-1)(q-t)p+(\alpha_1+\alpha_2)q^2
                -(-(\alpha_3+\alpha_4)t+\alpha_1+\alpha_2+\alpha_3)q-\alpha_4t ] ,
\end{multline*}
where $ q:=q_n, \hat{q}:=q_{n+1} $. From the definitions 
(\ref{VI_L14gDefn},\ref{VI_L14fDefn}) this result can be readily recast as
(\ref{dPV:a}). The second (\ref{dPV:b}) follows from a computation
for $ f_n+f_{n-1} $ using the shift operator $ L_{14} $.
\end{proof}

Making use of Lemma \ref{VI_L14seq}, together with elements of the Okamoto theory as 
detailed in \cite{FW_2002b}, and applying this to derive recurrences according to 
the strategy of \cite{FW_2003a}, gives the following recurrence scheme for 
$ \{\tau^{\rm VI}[N]\}_{N=0,1,2,\ldots} $.
\begin{proposition}\label{VI_L14dPV}
Let $\{g_N,f_N\}_{N=0,1,\dots}$, satisfy the \dPV system
\begin{align}
  g_{N+1}g_N 
  & = t{(f_N+N+1)(f_N+N+\mu+\omega) \over f_N(f_N-\mu-\bar{\omega})},
  \label{VI_L14Recur:a} \\
  f_N+f_{N-1} 
  & = \mu+\bar{\omega}+{N+2 \mu \over g_N-1}
      +{(N+2\omega_1)t \over g_N-t},
  \label{VI_L14Recur:b}
\end{align}
where $ t = e^{i\phi} $ subject to the initial conditions
\begin{gather*}
  g_0 = {q_0-t \over q_0-1}, 
  \\
  f_0 = {1 \over 1-t}\left[ (\mu+\omega)(q_0-1)+(\mu+\bar{\omega})(q_0-t)
                    - 2\omega_1{(q_0-t)(q_0-1) \over q_0} \right],
\end{gather*}
with
\begin{equation}
  q_0 = {\omega_1 \over \mu}
  { -i\dfrac{d}{d\phi}\log e^{i\mu\phi}T_1(e^{i\phi}) \over 
   \mu+\omega+i\dfrac{d}{d\phi}\log e^{i\mu\phi}T_1(e^{i\phi}) } .
\label{VI_L14Tinitial}
\end{equation}
Define $\{q_N,p_N\}_{N=0,1,\dots}$ in terms of $\{f_N,g_N\}_{N=0,1,\dots}$ by
\begin{gather}
  q_N = {g_N-t \over g_N-1}, \\
  p_N = {g_N-1 \over (1-t)g_N} \Big[
       (g_N-1)f_N-(\mu+\bar{\omega})g_N+(N+2\omega_1){(1-t)g_N \over g_N-t}
                 -N-\mu-\omega \Big].
\end{gather}
Then with $T_0(e^{i\phi}) = 1$ and $T_1(e^{i\phi})=w_0(e^{i\phi}) $ as given by
(\ref{VI_toepM},\ref{VI_toepM2}), $\{T_N\}_{N=2,3,\dots}$ is specified by 
the recurrence
\begin{multline}
  -(N+\mu+\omega)(N+\mu+\bar{\omega})
     {T_{N+1}T_{N-1} \over T_N^2 } \\
  = q_N(q_N-1)^2p^2_N+[(2\mu-N)q_N+N+2\omega_1](q_N-1)p_N-2\mu Nq_N-N(N+2\omega_1) .
\label{VI_L14Trecur}
\end{multline}
\end{proposition}
\begin{proof}
Let $ Y_n := L^{-1}_{14}K_n - K_n = K_{n+1}-K_n $. From Table 1 of \cite{FW_2002b} 
we have
\begin{equation*}
   Y_n = -{(t-1)q_n \over q_n-1} \left\{ (q_n-1)p_n+\alpha_0+\alpha_2-1
   + { (1-\alpha_0-\alpha_2)(\alpha_1+\alpha_2+\alpha_4) \over
        q_n(q_n-1)p_n+(\alpha_1+\alpha_2)q_n+\alpha_4 } \right\} .
\label{PVI_Y14}
\end{equation*}
Now consider
\begin{align*}
   t(t-1){d \over dt}\log{\tau_{n+1}\tau_{n-1} \over \tau_n^2}
  & = K_{n+1}+K_{n-1}-2K_n , \\
  & = Y_n-L_{14}Y_n .
\end{align*}
This latter difference, upon again consulting Table 1 of \cite{FW_2002b}, 
turns out to be
\begin{multline*}
  Y_n-L_{14}Y_n 
  = t(t-1){d \over dt}\log\bigg( q_n(q_n-1)^2p^2_n
  \\
    + [(\alpha_1+2\alpha_2)q_n+\alpha_4](q_n-1)p_n
       +\alpha_2[(\alpha_1+\alpha_2)q_n+\alpha_4] \bigg) .
\end{multline*}
After integrating both expressions and introducing an integration constant
(\ref{VI_L14Trecur}) follows.
\end{proof}

\begin{remark}
It is known from \cite{FW_2003a},\cite{FW_2002b} that the sequence of auxiliary 
variables $ \{g_n,f_n\}_{n=0,1,\ldots} $ defined by
\begin{gather}
  g_n := {q_n \over q_n - 1},
  \label{VI_L01gDefn} \\
  f_n := q_n(q_n-1)p_n + (1-\alpha_2-\alpha_4)(q_n-1)-\alpha_3q_n
                    - \alpha_0{q_n(q_n-1) \over q_n-t},
  \label{VI_L01fDefn}
\end{gather}
generated by the shift operator $ L^{-1}_{01} $ satisfy the \dPV equations
\begin{align}
  g_{n+1}g_n 
  & = {t\over t-1}{(f_n+1-\alpha_2)(f_n+1-\alpha_2-\alpha_4) \over f_n(f_n+\alpha_3)},
  \label{VI_L01gRecur} \\
  f_n+f_{n-1} & = -\alpha_3 + {\alpha_1 \over g_n-1} +
                  {\alpha_0 t \over t(g_n-1)-g_n}.
  \label{VI_L01fRecur}
\end{align}
In fact the two systems of recurrences (\ref{VI_L01gRecur},\ref{VI_L01fRecur}) and 
(\ref{dPV:a},\ref{dPV:b}) are related by an element of the $ S_4 $ subgroup
of the $ W_a(F_4) $ transformations, namely the generator $ x^3 $ \cite{Ok_1987a}.
This has the action 
\begin{equation}
   x^3: \alpha_0 \leftrightarrow \alpha_4, t \mapsto {t \over t-1},
        q \mapsto {t-q \over t-1}, p \mapsto -(t-1)p ,
\end{equation}
and when applying these transformations to (\ref{VI_L01gDefn}), (\ref{VI_L01fDefn}),
(\ref{VI_L01gRecur}), (\ref{VI_L01fRecur}) we recover (\ref{VI_L14gDefn}), 
(\ref{VI_L14fDefn}), (\ref{dPV:a}), (\ref{dPV:b}) respectively.
\end{remark}

\section{Applications to Physical Models}
\label{Modelsection}
\setcounter{equation}{0}

\subsection{Random Matrix Averages}

A specialisation of the above results with great interest in the application of 
random matrices \cite{KS_2000b},\cite{KM_2004} is the quantity
\begin{equation}
   F^{\rm CUE}_N(u;\mu) := 
   \Big\langle \prod^{N}_{l=1}|u+z_l|^{2\mu} \Big\rangle_{{\rm CUE}_N} .
\label{PVI_CUE}
\end{equation}
This has the interpretation as the average of the $2\mu$-th power, or equivalently
the $2\mu$-th moment, of the absolute value of the characteristic polynomial for 
the CUE. In the case $ |u|=1 $ (\ref{PVI_CUE}) is independent of $ u $ and has 
the well-known (see e.g. \cite{BF_1997}) gamma function evaluation
\begin{equation}
 \Big\langle \prod^{N}_{l=1}|u+z_l|^{2\mu} \Big\rangle_{{\rm CUE}_N}\Big|_{u=e^{i\phi}}
  = \Big\langle \prod^{N}_{l=1}|1+z_l|^{2\mu} \Big\rangle_{{\rm CUE}_N}
  = \prod^{N-1}_{j=0}{j!\Gamma(j+1+2\mu) \over \Gamma^2(j+1+\mu)} ,
\end{equation}
where for convergence of the integral $ \Re(\mu) > -\half $. For $ |u|<1 $ we see 
by an appropriate change of variables that
\begin{align}                                       
    \Big\langle \prod^{N}_{l=1}|u+z_{l}|^{2\mu} \Big\rangle_{{\rm CUE}_N}
  & = \Big\langle \prod^{N}_{l=1}(1+|u|^2z_{l})^{\mu}(1+1/z_{l})^{\mu} \Big\rangle_{{\rm CUE}_N} ,
  \nonumber \\
  & = {}^{\vphantom{(1)}}_{2}F^{(1)}_{1}(-\mu,-\mu;N;t_1,\ldots,t_N)|_{t_1=\ldots =t_N=|u|^2},
\label{CUEp_genH}
\end{align}
where the second equality follows from (\ref{VI_2F1}). 
For $ |u|>1 $ we can use the simple functional equation
\begin{equation}                                       
    \Big\langle \prod^{N}_{l=1}|u+z_{l}|^{2\mu} \Big\rangle_{{\rm CUE}_N}
  = |u|^{2\mu N}\Big\langle \prod^{N}_{l=1}|{1\over u}+z_{l}|^{2\mu} \Big\rangle_{{\rm CUE}_N},
\end{equation}
to relate this case back to the case $ |u|<1 $.

The weight in the first equality of (\ref{CUEp_genH}) is a special case of 
(\ref{VI_wgt}). In terms of the parameters of the form (\ref{VI_wgt}) we observe 
that $ \xi = 0 $, $ 2\mu \mapsto \mu $, $ \omega = \bar{\omega} = \mu/2 $, 
i.e. $ \omega_2 = 0 $ and $ t = |u|^2 $. The trigonometric moments are 
\begin{align}
   w_{-n} & =
   {\Gamma(\mu+1) \over n!\Gamma(\mu+1-n)}{}_2F_1(-\mu,-\mu+n;n+1;|u|^2)
   \qquad n \in \mathbb{Z}_{\geq 0}, \\
   w_{n}  & = |u|^{2n}w_{-n} \qquad n \in \mathbb{Z}_{\geq 0} .
\end{align}
The results of Section \ref{PVIsection} then allow (\ref{PVI_CUE})
to be computed by a recurrence involving the corresponding reflection coefficients.

\begin{corollary}
The general moments of the characteristic polynomial $ |\det(u+U)| $ for arbitrary 
exponent $ 2\mu $ with respect to the finite CUE ensemble $ U \in U(N) $ of rank $ N $ 
is given by the system of recurrences
\begin{equation}
  {F^{\rm CUE}_{N+1}F^{\rm CUE}_{N-1} \over (F^{\rm CUE}_{N})^2} = 1-|u|^{2N}r^2_N ,
\end{equation}
with initial values
\begin{equation}
  F^{\rm CUE}_{0} = 1, \qquad F^{\rm CUE}_{1} = {}_2F_1(-\mu,-\mu;1;|u|^2) , 
\end{equation}
and 
\begin{multline}
   2|u|^{2N}r_{N}r_{N-1}-|u|^2-1
   \\
   = {1-|u|^{2N}r^2_{N} \over r_{N}}
       \left[ (N+1+\mu)|u|^2r_{N+1} + (N-1+\mu)r_{N-1} \right]
   \\
   - {1-|u|^{2(N-1)}r^2_{N-1} \over r_{N-1}}
       \left[ (N+\mu)|u|^2r_{N} + (N-2+\mu)r_{N-2} \right],
   \label{CUEp_rRecur}
\end{multline}
subject to the initial values
\begin{equation}
   r_{0} = 1, \quad
   r_{1} = -\mu{{}_2F_1(-\mu,-\mu+1;2;|u|^2) \over {}_2F_1(-\mu,-\mu;1;|u|^2)} .
\end{equation}
\end{corollary}

\begin{proof}
From either (\ref{VI_2ndRR}), (\ref{VI_lRecur}) or (\ref{VI_2+1rRecur:a}) and 
the fact that $ \bar{r}_1 = |u|^2r_1 $ we can repeat the arguments of Corollary
\ref{realRC} to deduce that $ \bar{r}_N = |u|^{2N}r_N $ for $ N \geq 0 $.
The recurrence relation follows simply from the specialisation of 
(\ref{VI_rRecur:a}) and the initial conditions from the $ N=1 $ case.
\end{proof}
  
Another spectral statistic of fundamental importance in random matrix theory
is the generating function $ E^{\rm CUE}_N((\pi-\phi,\pi);\xi) $ for the 
probabilities $ E^{\rm CUE}_N(k;(\pi-\phi,\pi)) $ that exactly $ k $ eigenvalues
lie in the interval $ (\pi-\phi,\pi) $. This is specified by
\begin{multline}
 E^{\rm CUE}_N((\pi-\phi,\pi);\xi) 
 \\ 
 := {1 \over C_N}
 \left( \int^{\pi}_{-\pi} - \xi\int^{\pi}_{\pi-\phi} \right) d\theta_1 \ldots
 \left( \int^{\pi}_{-\pi} - \xi\int^{\pi}_{\pi-\phi} \right) d\theta_N
 \prod_{1 \leq j < k \leq N} |e^{i\theta_j}-e^{i\theta_k}|^2 ,
\label{PVI_CUE:b}
\end{multline}
where $ C_N = (2\pi)^N N! $, and thus corresponds to the 
special case $ \mu = \omega = \bar{\omega} = 0 $, $ t=e^{i\phi} $ of (\ref{VI_wgt}).
We remark that for these parameters the underlying Toeplitz matrix elements, 
given for general parameters by (\ref{VI_toepM}) or (\ref{VI_toepM2}), have the
elementary form
\begin{equation}
   w_n = \delta_{n,0} + {\xi \over 2\pi i}(-1)^{n+1}{t^n-1 \over n} ,
\end{equation}

A recurrence scheme for the generating function $ E^{\rm CUE}_N $, deduced as a
corollary of Proposition \ref{VI_L01dPV}, has been presented in \cite{FW_2003a}. 
Here we use recurrences found herein for $ r_N $, $ \bar{r}_N $, together with the 
fact that for (\ref{PVI_CUE:b}) one has $ r_N = t^{-N}\bar{r}_N $, to replace
the role of the coupled recurrences from \cite{FW_2003a} by a single recurrence.
\begin{corollary}\label{CUE_recur}
The generating function for the probability of finding exactly $ k $ eigenvalues 
$ z = e^{i\theta} $ from the ensemble of random $ N \times N $ unitary matrices
within the sector of the unit circle $ \theta \in (\pi-\phi,\pi] $ 
is given by the following system of recurrences in the rank of the ensemble $ N $,
\begin{equation}
   {E^{\rm CUE}_{N+1}E^{\rm CUE}_{N-1} \over (E^{\rm CUE}_{N})^2}
  = 1 - x^2_N ,
\end{equation}
where the initial values are 
\begin{equation}
  E^{\rm CUE}_{0} = 1, \quad E^{\rm CUE}_{1} = 1 - {\xi \over 2\pi}\phi ,
\end{equation}
and the auxiliary variables $ x_N $ are determined by the quasi-linear third order
recurrence relation
\begin{multline}
   2x_{N}x_{N-1} - 2\cos{\phi \over 2}
  = {1-x^2_N \over x_N}\left[ (N+1)x_{N+1}+(N-1)x_{N-1} \right] \\
    - {1-x^2_{N-1} \over x_{N-1}}\left[ Nx_{N}+(N-2)x_{N-2} \right] ,
\end{multline}
or the quadratic second order recurrence relation
\begin{multline}
 (1-x^2_N)^2\left[ (N+1)^2x^2_{N+1}+(N-1)^2x^2_{N-1} \right]
   + 2(N^2-1)(1-x^4_N)x_{N+1}x_{N-1} \\
 + 4N\cos{\phi \over 2}x_N(1-x^2_N)\left[ (N+1)x_{N+1}+(N-1)x_{N-1} \right]
 \\ + 4N^2x^2_N\left[ \cos^2{\phi \over 2}-x^2_N \right] = 0,
\end{multline}
along with the initial values
\begin{equation}
   x_{-1} = 0, \quad x_{0} = 1, \quad
   x_{1} = -{\xi \over \pi}{\sin\dfrac{\phi}{2} \over 
                            1 - \dfrac{\xi}{2\pi}\phi } .
\end{equation} 
\begin{proof}
The first recurrence relation follows directly from the general recurrence 
(\ref{VI_rRecur:a}) and Corollary \ref{realRC} whilst the second follows from 
(\ref{VI_1+1rRecur:a}).
\end{proof}
\end{corollary}

\subsection{2-D Ising Model}

Consider the two-dimensional Ising model with dimensionless nearest neighbour
couplings equal to $ K_1 $ and $ K_2 $ in the $ x $ and $ y $ directions 
respectively (see e.g. \cite{Ba_1982}). Let $ \sigma_{0,0} $ and $ \sigma_{N,N} $
denote the values of the spins at the lattice sites $ (0,0) $ and $ (N,N) $
respectively. For the infinite lattice an unpublished result of Onsager
(see \cite{McCW_1973}) gives that the diagonal spin-spin correlation
$ \langle \sigma_{0,0}\sigma_{N,N} \rangle $ has the Toeplitz form
\begin{equation}
   \langle \sigma_{0,0}\sigma_{N,N} \rangle = 
   \det (a_{i-j}(k))_{1 \leq i,j \leq N} ,
\label{IM_ssDiag}
\end{equation}
where
\begin{equation}
   a_{p}(k) := 
   \frac{1}{2\pi} \int^{2\pi}_{0} d\theta e^{-ip\theta}
   \left[\frac{1-(1/k)e^{-i\theta}}{1-(1/k)e^{i\theta}}\right]^{1/2},
\label{IM_ssSymbol}
\end{equation}
$ k = \sinh2K_{1}\sinh2K_{2} $. The analytic structure is different depending on
$ k > 1 $ (low temperature phase) or $ k < 1 $ (high temperature phase). 

Jimbo and Miwa \cite{JM_1980} identified (\ref{IM_ssDiag}) as the $ \tau$-function
associated with a monodromy preserving deformation of a linear system, which in
turn is associated with \PVI. In a more recent work \cite{FW_2002b} the present 
authors have identified (\ref{IM_ssDiag}) as a $ \tau$-function in the Okamoto
theory of \PVI \cite{Ok_1987a}. These identifications have the consequence of
allowing (\ref{IM_ssDiag}) to be characterised in terms of a solution of the 
$ \sigma$-form of the Painlev\'e VI equation, when regarded as a function of
$ t:=k^2 $.

It is our objective in this subsection to use the recurrences of Subsection 3.1,
appropriately specialised, to derive a $ N$-recurrence for (\ref{IM_ssDiag}).
Before doing so, we remark that a recent result of Borodin \cite{Bo_2001} can
also be used for the same purpose. The latter recurrence applies to all Toeplitz 
determinants
\begin{equation}
   q^{(z,z',\xi)}_n := (1-\xi)^{zz'}\det[g_{i-j}]_{i,j=1,\ldots,n} ,
\label{}
\end{equation}
where
\begin{equation}
   g_p = {1\over 2\pi} \int^{2\pi}_{0} d\theta e^{-ip\theta} 
                       \left(1-\sqrt{\xi}e^{i\theta}\right)^{z}
                       \left(1-\sqrt{\xi}e^{-i\theta}\right)^{z'} .
\label{}
\end{equation}
As previously noted in \cite{FW_2002b}, comparison with (\ref{IM_ssDiag}) and 
(\ref{IM_ssSymbol}) shows
\begin{equation}
   \langle \sigma_{0,0}\sigma_{N,N} \rangle 
  = (1-k^{-2})^{1/4}q^{(-1/2,1/2,1/k^2)}_N .
\label{}
\end{equation}

By regarding the Fourier integral in (\ref{IM_ssSymbol}) as a contour integral,
and changing the contour of integration, the corresponding weight function can
be chosen to be equal to \cite{FW_2002b}
\begin{align}
   z^{1/4}|1+z|^{-1/2}(1+k^{-2}z)^{1/2}
   & = z^{1/2}(1+z)^{-1/2}(1+k^{-2}z)^{1/2}, \quad 1/k^2 < 1 ,
   \label{IM_wgt:a} \\
   k^{-1}z^{-3/4}|1+z|^{-1/2}(1+k^{2}z)^{1/2}
   & = k^{-1}z^{-1/2}(1+z)^{-1/2}(1+k^{2}z)^{1/2}, \quad k^2 < 1 .
   \label{IM_wgt:b}
\end{align}
The first, multiplied by $ k^{1/2} $, is the case 
\begin{equation}
  \xi = 0, \mu = \frac{1}{4}, \omega_1 = -\frac{1}{4}, \omega_2 = \frac{i}{2},  t = 1/k^2 , 
\label{IM_param:a}
\end{equation}
of (\ref{VI_wgt}), while the second multiplied by $ k^{1/2} $ is the case 
\begin{equation}
  \xi = 0, \mu = \frac{1}{4}, \omega_1 = -\frac{1}{4}, \omega_2 = -\frac{i}{2},  t = k^2 .
\label{IM_param:b}
\end{equation}
Substituting these values in (\ref{VI_toepM}) gives Gauss hypergeometric evaluations
for the matrix elements (\ref{IM_ssSymbol}). After an appropriate limiting 
procedure to account for factors which otherwise would be singular, one obtains
the well known fact that the Toeplitz elements in the low temperature regime are 
given by
\begin{align}
  w_{-n} & = {(-1)^n \over \pi}{\Gamma(n+\half)\Gamma(\half) \over \Gamma(n+1)}
             {}_2F_1(-\half,n+\half;n+1;k^{-2}), \quad n \geq 0,
         \label{IM_Toep:a}\\
  w_{n}  & = {(-1)^{n+1}k^{-2n} \over \pi}{\Gamma(n-\half)\Gamma(\thalf) \over \Gamma(n+1)}
             {}_2F_1(\half,n-\half;n+1;k^{-2}), \quad n > 0,
         \label{IM_Toep:b}
\end{align}
whilst those in the high temperature regime are
\begin{align}
  w_{-n} & = {(-1)^nk^{2n+1} \over \pi}{\Gamma(n+\half)\Gamma(\thalf) \over \Gamma(n+2)}
             {}_2F_1(\half,n+\half;n+2;k^{2}), \quad n \geq 0,
         \label{IM_Toep:c}\\
  w_{n}  & = {(-1)^{n-1} \over \pi k}{\Gamma(n-\half)\Gamma(\half) \over \Gamma(n)}
             {}_2F_1(-\half,n-\half;n;k^{2}), \quad n > 0.
         \label{IM_Toep:d}
\end{align}

The results of Subsection 3.1 provide the following recurrence scheme.
\begin{corollary}\label{IM_recur}
The diagonal correlation function for the Ising model valid in both the low and
high temperature phases (with $ k \mapsto 1/k $ in the latter case) is determined by
\begin{equation}
 \frac{\langle \sigma_{0,0}\sigma_{N+1,N+1} \rangle\langle \sigma_{0,0}\sigma_{N-1,N-1} \rangle}
      {\langle \sigma_{0,0}\sigma_{N,N} \rangle^2}
  = 1-r_{N}\bar{r}_{N},
\end{equation}
along with the quasi-linear $ 2/1 $ 
\begin{multline}
   (2N+3)k^{-2}(1-r_{N}\bar{r}_{N})r_{N+1}
     + 2N\left[ k^{-2}+1-(2N-1)k^{-2}r_{N}\bar{r}_{N-1} \right]r_{N} \\
     + (2N-3)\left[ (2N-1)r_{N}\bar{r}_{N}+1 \right]r_{N-1} = 0,
   \label{ising_rRecur:a}
\end{multline}
and $ 1/2 $ recurrence relations 
\begin{multline}
   (2N+1)(1-r_{N}\bar{r}_{N})\bar{r}_{N+1}
     + 2N\left[ (2N-3)\bar{r}_{N}r_{N-1}+k^{-2}+1 \right]\bar{r}_{N} \\
     + (2N-1)k^{-2}\left[ -(2N+1)r_{N}\bar{r}_{N}+1 \right]\bar{r}_{N-1} = 0,
   \label{ising_rRecur:b}
\end{multline}
subject to initial conditions for the low temperature regime
\begin{equation}                                       
   r_{0} = \bar{r}_{0} = 1, \;
   r_{1} = {2-k^2 \over 3}+{k^2-1 \over 3}{{\rm K}(k^{-1}) \over {\rm E}(k^{-1})},
   \;
   \bar{r}_{1} = -1+{k^2-1 \over k^2}{{\rm K}(k^{-1}) \over {\rm E}(k^{-1})},
\end{equation}
or to the initial conditions for the high temperature regime given by
\begin{equation}                                       
   r_{0} = \bar{r}_{0} = 1, \;
   r_{1} = {1 \over 3}\left\{
            {2 \over k^2}-{{\rm E}(k) \over (k^2-1){\rm K}(k)+{\rm E}(k)} \right\},
   \;
   \bar{r}_{1} = -{k^2{\rm E}(k) \over (k^2-1){\rm K}(k)+{\rm E}(k)},
\end{equation}
where $ {\rm K}(k), {\rm E}(k) $ are the complete elliptic integrals of the first
and second kind respectively.
\end{corollary}

\begin{proof}
The equations (\ref{ising_rRecur:a},\ref{ising_rRecur:b}) follow from 
(\ref{VI_2+1rRecur:a}) and its "conjugate" upon specialisation of the parameters
as required by (\ref{IM_param:a}) and (\ref{IM_param:b}). The initial conditions 
follow from (\ref{VI_rInitial}), (\ref{IM_Toep:a})-(\ref{IM_Toep:d}) 
and the relationship between the Gauss hypergeometric functions therein and the 
complete elliptic integrals.
\end{proof}

A consequence of the results of Subsection 3.3 is the following result relating
the quantities of Corollary \ref{IM_recur} to the generalised hypergeometric 
function $ {}_2F^{(1)}_1 $.
\begin{corollary}
In the low temperature phase the diagonal correlation function is given by
\begin{equation}                                       
    \langle \sigma_{0,0}\sigma_{N,N} \rangle =
  {}^{\vphantom{(1)}}_{2}F^{(1)}_{1}(-\half,\half;N;t_1,\ldots,t_N)|_{t_1=\ldots =t_N=1/k^2},
\label{ising_genH:a}
\end{equation}
whilst the reflection coefficients are given by
\begin{align}
   r_{N} 
  & = (-1)^{N}{(-\half)_{N} \over N!}
  { {}^{\vphantom{(1)}}_{2}F^{(1)}_{1}(-\half,\thalf;N+1;t_1,\ldots,t_N)
      \over 
    {}^{\vphantom{(1)}}_{2}F^{(1)}_{1}(-\half,\half;N;t_1,\ldots,t_N) 
  }\Bigg|_{t_1=\ldots =t_N=1/k^2},
  \label{ising_genH:c}\\
   \bar{r}_{N} 
  & = (-1)^{N}{(N-1)! \over (\half)_{N}}
  { \lim_{\epsilon \to 0}\epsilon
    {}^{\vphantom{(1)}}_{2}F^{(1)}_{1}(-\half,-\half;N-1+\epsilon;t_1,\ldots,t_N)
      \over 
    {}^{\vphantom{(1)}}_{2}F^{(1)}_{1}(-\half,\half;N;t_1,\ldots,t_N)
  }\Bigg|_{t_1=\ldots =t_N=1/k^2}.
  \label{ising_genH:d}
\end{align}
In the high temperature phase the diagonal correlation function is
\begin{equation}                                       
    \langle \sigma_{0,0}\sigma_{N,N} \rangle =
   \frac{(2N-1)!!}{2^NN!}k^N
  {}^{\vphantom{(1)}}_{2}F^{(1)}_{1}(\half,\half;N+1;t_1,\ldots,t_N)|_{t_1=\ldots =t_N=k^2},
\label{ising_genH:b}
\end{equation}
and the reflection coefficients are given by
\begin{align}
   r_{N} 
  & = (-1)^{N}{(-\half)_{N} \over (N+1)!}
  { {}^{\vphantom{(1)}}_{2}F^{(1)}_{1}(\half,\thalf;N+2;t_1,\ldots,t_N)
      \over 
    {}^{\vphantom{(1)}}_{2}F^{(1)}_{1}(\half,\half;N+1;t_1,\ldots,t_N) 
  }\Bigg|_{t_1=\ldots =t_N=k^2},
  \label{ising_genH:e}\\
   \bar{r}_{N} 
  & = (-1)^{N}{N! \over (\half)_{N}}
  { {}^{\vphantom{(1)}}_{2}F^{(1)}_{1}(\half,-\half;N;t_1,\ldots,t_N)
      \over 
    {}^{\vphantom{(1)}}_{2}F^{(1)}_{1}(\half,\half;N+1;t_1,\ldots,t_N)
  }\Bigg|_{t_1=\ldots =t_N=k^2}.
  \label{ising_genH:f}
\end{align}
\end{corollary}

\begin{proof}
The evaluations (\ref{ising_genH:a}) and (\ref{ising_genH:b}) follow by specialising
the parameters in (\ref{VI_2F1}) as required by (\ref{IM_param:a}) and 
(\ref{IM_param:b}) respectively. The evaluations (\ref{ising_genH:c}), 
(\ref{ising_genH:d}) in the low temperature phase follow from (\ref{VI_genH:a}), 
(\ref{VI_genH:b}) by specialising the parameters as required by (\ref{IM_param:a}).
Some care needs to be taken with $  \bar{r}_{N} $ because $ -\mu+\bar{\omega} = 0 $. 
Applying a limiting process leads to
\begin{multline}
\lim_{\epsilon \to 0}\epsilon
    {}^{\vphantom{(1)}}_{2}F^{(1)}_{1}(-\half,-\half;N-1+\epsilon;t_1,\ldots,t_N) \\
  = \sum_{\kappa:l(\kappa) = N}
    {([-\half]^{(1)}_{\kappa})^2 \over [N]^{(1)}_{\kappa}}
    {\prod^{N}_{j=1}(N-j+\kappa_{j}) \over (N-1)!}
    {s_{\kappa}(t_1,\ldots,t_N) \over h_{\kappa}} ,
\end{multline}
in which only those terms with lengths $ l(\kappa) = N $ contribute to the sum.
The high temperature expressions follow from the low temperature ones through 
the transformation $ \mu \leftrightarrow \omega_1 $.
\end{proof}

It is of interest to note that as $ N $ grows more of the leading order terms in 
the expansion of (\ref{ising_genH:a}) become independent of $ N $, and the
following limit becomes explicit
\begin{equation}                                       
  \lim_{N \to \infty} \langle \sigma_{0,0}\sigma_{N,N} \rangle
   = (1-k^{-2})^{1/4}.
\label{}
\end{equation}

At zero temperature, $ k = \infty $, the solutions simplify to
\begin{equation}                                       
   r_{N} = (-1)^{N}{(-\half)_{N} \over N!}, \quad
   \bar{r}_{N} = 0 \; (N \geq 1), \quad 
    \langle \sigma_{0,0}\sigma_{N,N} \rangle = 1,
\end{equation}
whilst at the critical point, $ k = 1 $, we have the simple solutions
\begin{gather}                                       
   r_{N} = {(-1)^{N-1} \over (2N+1)(2N-1)}, \quad
   \bar{r}_{N} = (-1)^{N}, \quad l_{N} = {N \over 2N+1}, \\
    \langle \sigma_{0,0}\sigma_{N,N} \rangle = 
    \prod^{N}_{j=1}{\Gamma^2(j) \over \Gamma(j+\half)\Gamma(j-\half)} .
\end{gather}
At infinite temperature they become
\begin{equation}                                       
   r_{N} = (-1)^{N}{(-\half)_{N} \over (N+1)!}, \quad
   \bar{r}_{N} = (-1)^{N}{ N! \over (\half)_{N}}, \quad 
    \langle \sigma_{0,0}\sigma_{N,N} \rangle = 0 \; (N \geq 1) .
\end{equation}
These are all in agreement with the known results \cite{McCW_1973}.

\section*{Acknowledgments}

This research has been supported by the Australian Research Council. NSW appreciates
the generosity of Will Orrick in supplying expansions of the Toeplitz determinants for
the diagonal correlations of the Ising model and the assistance of Paul Leopardi in 
calculating gap probabilities for the CUE.

\bibliographystyle{amsplain}
\bibliography{moment,nonlinear,random_matrices}

\def\cprime{$'$} \def\cprime{$'$} \def\cydot{\leavevmode\raise.4ex\hbox{.}}
  \def\cprime{$'$} \def\cprime{$'$} \def\cprime{$'$} \def\cprime{$'$}
  \def\cprime{$'$} \def\cprime{$'$} \def\cprime{$'$} \def\cprime{$'$}
  \def\cprime{$'$} \def\cdprime{$''$} \def\cydot{\leavevmode\raise.4ex\hbox{.}}
  \def\cydot{\leavevmode\raise.4ex\hbox{.}} \def\cprime{$'$}
  \def\cydot{\leavevmode\raise.4ex\hbox{.}} \def\cprime{$'$} \def\cprime{$'$}
  \def\cprime{$'$} \def\cprime{$'$} \def\cprime{$'$} \def\cprime{$'$}
  \def\cprime{$'$} \def\cprime{$'$} \def\cprime{$'$} \def\cprime{$'$}
  \def\cprime{$'$} \def\cprime{$'$} \def\cydot{\leavevmode\raise.4ex\hbox{.}}
  \def\cprime{$'$} \def\cprime{$'$} \def\cprime{$'$} \def\cprime{$'$}
  \def\cprime{$'$} \def\cprime{$'$} \def\cprime{$'$} \def\cprime{$'$}
  \def\cprime{$'$} \def\cprime{$'$} \def\cprime{$'$} \def\cprime{$'$}
  \def\cprime{$'$} \def\cprime{$'$} \def\cprime{$'$} \def\cprime{$'$}
  \def\cprime{$'$} \def\cprime{$'$} \def\cprime{$'$} \def\cprime{$'$}
  \def\cprime{$'$} \def\cprime{$'$} \def\cprime{$'$} \def\cprime{$'$}
  \def\cprime{$'$} \def\cprime{$'$} \def\cydot{\leavevmode\raise.4ex\hbox{.}}
  \def\cprime{$'$} \def\cprime{$'$} \def\cprime{$'$} \def\cprime{$'$}
  \def\cprime{$'$} \def\cprime{$'$} \def\cprime{$'$} \def\cprime{$'$}
  \def\cprime{$'$} \def\cprime{$'$} \def\cprime{$'$} \def\cprime{$'$}
  \def\cprime{$'$} \def\cprime{$'$} \def\cprime{$'$}
  \def\cydot{\leavevmode\raise.4ex\hbox{.}} \def\cprime{$'$} \def\cprime{$'$}
  \def\cprime{$'$} \def\cprime{$'$} \def\cprime{$'$} \def\cprime{$'$}
  \def\cprime{$'$} \def\cprime{$'$} \def\cprime{$'$} \def\cprime{$'$}
  \def\cprime{$'$} \def\cprime{$'$} \def\cprime{$'$} \def\cprime{$'$}
  \def\cprime{$'$} \def\cprime{$'$} \def\cprime{$'$} \def\cprime{$'$}
  \def\cprime{$'$} \def\cprime{$'$} \def\cprime{$'$} \def\cprime{$'$}
  \def\cprime{$'$} \def\cprime{$'$} \def\cprime{$'$} \def\cprime{$'$}
  \def\cprime{$'$} \def\cprime{$'$} \def\cprime{$'$} \def\cprime{$'$}
  \def\cprime{$'$} \def\cprime{$'$} \def\cydot{\leavevmode\raise.4ex\hbox{.}}
  \def\cydot{\leavevmode\raise.4ex\hbox{.}}
  \def\cydot{\leavevmode\raise.4ex\hbox{.}}
  \def\cydot{\leavevmode\raise.4ex\hbox{.}}
  \def\cydot{\leavevmode\raise.4ex\hbox{.}}
  \def\cydot{\leavevmode\raise.4ex\hbox{.}}
  \def\cydot{\leavevmode\raise.4ex\hbox{.}}
  \def\cydot{\leavevmode\raise.4ex\hbox{.}} \def\cprime{$'$} \def\cprime{$'$}
\providecommand{\bysame}{\leavevmode\hbox to3em{\hrulefill}\thinspace}
\providecommand{\MR}{\relax\ifhmode\unskip\space\fi MR }
\providecommand{\MRhref}[2]{%
  \href{http://www.ams.org/mathscinet-getitem?mr=#1}{#2}
}
\providecommand{\href}[2]{#2}
\begin{thebibliography}{10}

\bibitem{AvM_2002}
M.~Adler and P.~van Moerbeke, \emph{Recursion relations for unitary integrals,
  combinatorics and the {T}oeplitz lattice}, Comm. Math. Phys. \textbf{237}
  (2003), no.~3, 397--440. \MR{1 993 333}

\bibitem{BR_2001}
J.~Baik and E.~M. Rains, \emph{Algebraic aspects of increasing subsequences},
  Duke Math. J. \textbf{109} (2001), no.~1, 1--65. \MR{1 844 203}

\bibitem{BF_1997}
T.~H. Baker and P.~J. Forrester, \emph{The {C}alogero-{S}utherland model and
  generalized classical polynomials}, Comm. Math. Phys. \textbf{188} (1997),
  no.~1, 175--216. \MR{99c:33012}

\bibitem{Ba_1982}
R.~J. Baxter, \emph{Exactly {S}olved {M}odels in {S}tatistical {M}echanics},
  Academic Press, London, 1982.

\bibitem{BO_2000}
A.~Borodin and G.~Olshanski, \emph{Distributions on partitions, point
  processes, and the hypergeometric kernel}, Comm. Math. Phys. \textbf{211}
  (2000), no.~2, 335--358. \MR{2001k:33031}

\bibitem{Bo_2001}
Alexei Borodin, \emph{Discrete gap probabilities and discrete {P}ainlev\'e
  equations}, Duke Math. J. \textbf{117} (2003), no.~3, 489--542. \MR{1 979
  052}

\bibitem{BB_2002}
Alexei Borodin and Dmitriy Boyarchenko, \emph{Distribution of the first
  particle in discrete orthogonal polynomial ensembles}, Comm. Math. Phys.
  \textbf{234} (2003), no.~2, 287--338. \MR{1 962 463}

\bibitem{Fo_1992}
P.~J. Forrester, \emph{Selberg correlation integrals and the {$1/r\sp 2$}
  quantum many-body system}, Nuclear Phys. B \textbf{388} (1992), no.~3,
  671--699. \MR{MR1201273 (94e:33030)}

\bibitem{FFGW_2002a}
P.~J. Forrester, N.~E. Frankel, T.~Garoni, and N.~S. Witte, \emph{Painlev\'e
  transcendent evaluations of finite system density matrices for 1d
  impenetrable bosons}, Comm. Math. Phys. \textbf{238} (2003), no.~1-2,
  257--285.

\bibitem{FW_2003a}
P.~J. Forrester and N.~S. Witte, \emph{Discrete {P}ainlev\'e equations and
  random matrix averages}, Nonl. \textbf{16} (2003), no.~6, 1919--1944.

\bibitem{FW_2002b}
\bysame, \emph{{A}pplication of the $\tau$-function theory of {P}ainlev{\'e}
  equations to random matrices: {PVI}, the {JUE}, {CyUE}, {cJUE} and scaled
  limits}, Nagoya Math. J. \textbf{174} (2004), 29--114.

\bibitem{FW_2004a}
\bysame, \emph{Bi-orthogonal {P}olynomials on the {U}nit {C}ircle, regular
  semi-classical {W}eights and {I}ntegrable {S}ystems}, 2004.

\bibitem{FW_2003b}
\bysame, \emph{Discrete {P}ainlev\'e {E}quations, {O}rthogonal {P}olynomials on
  the {U}nit {C}ircle and $n$-recurrences for averages over {$U(N)$} --
  {\PIIIa} and {\PV} $\tau$-functions}, Int. Math. Res. Not. \textbf{2004}
  (2004), no.~4, 159--183.

\bibitem{Fr_1976}
G{\'e}za Freud, \emph{On the coefficients in the recursion formulae of
  orthogonal polynomials}, Proc. Roy. Irish Acad. Sect. A \textbf{76} (1976),
  no.~1, 1--6. \MR{54 \#7913}

\bibitem{JM_1980}
M.~Jimbo and T.~Miwa, \emph{Studies on holonomic quantum fields. {X}{V}{I}{I}},
  Proc. Japan Acad. Ser. A Math. Sci. \textbf{56} (1980), no.~9, 405--410.
  \MR{85h:82016}

\bibitem{Ka_1993}
J.~Kaneko, \emph{Selberg integrals and hypergeometric functions associated with
  {J}ack polynomials}, SIAM J. Math. Anal. \textbf{24} (1993), no.~4,
  1086--1110. \MR{94h:33010}

\bibitem{KM_2004}
J.~P. Keating and F.~Mezzadri, \emph{{R}andom {M}atrix {T}heory and
  {E}ntanglement in {Q}uantum {S}pin {C}hains}, {\tt arXiv:quant-ph/0407047}.

\bibitem{KS_2000b}
J.~P. Keating and N.~C. Snaith, \emph{Random matrix theory and
  {$\zeta(1/2+it)$}}, Comm. Math. Phys. \textbf{214} (2000), no.~1, 57--89.
  \MR{2002c:11107}

\bibitem{Kl_1933}
F.~Klein, \emph{{V}orlesungen {\"u}ber die {H}ypergeometrische {F}unction},
  Verlag von {J}ulius {S}pringer, Inc., Berlin, 1933, Gehalten an der
  {U}niversit{\"a}t {G}{\"o}ttingen in {W}intersemester 1893{/}94.

\bibitem{Ma_2000}
A.~P. Magnus, \emph{{MAPA3072A} {S}pecial topics in approximation theory
  1999-2000: {S}emi-classical orthogonal polynomials on the unit circle}, {\tt
  http://www.math.ucl.ac.be/\~{}magnus/}.

\bibitem{McCW_1973}
B.~McCoy and T.~T. Wu, \emph{The {T}wo-{D}imensional {I}sing {M}odel}, Harvard
  University Press, Harvard, 1973.

\bibitem{Ok_1987a}
K.~Okamoto, \emph{Studies on the {P}ainlev\'e equations. {I}. {S}ixth
  {P}ainlev\'e equation ${P}\sb {{\rm {V}{I}}}$}, Ann. Mat. Pura Appl. (4)
  \textbf{146} (1987), 337--381. \MR{88m:58062}

\bibitem{Sa_2001}
H.~Sakai, \emph{Rational surfaces associated with affine root systems and
  geometry of the {P}ainlev\'e equations}, Comm. Math. Phys. \textbf{220}
  (2001), no.~1, 165--229. \MR{1 882 403}

\bibitem{Ya_1992}
Z.~M. Yan, \emph{A class of generalized hypergeometric functions in several
  variables}, Canad. J. Math. \textbf{44} (1992), no.~6, 1317--1338.
  \MR{94c:33026}

\end{thebibliography}

\end{document}